\documentclass{article}
\usepackage{mathptmx,amsmath,amssymb,pgfplots,graphicx,rotating,hyperref}
\usepackage{float}

\usepackage{color}

\newcommand{\fracp}[2]{\frac{\partial#1}{\partial#2}}
\usetikzlibrary{decorations.text}
\oddsidemargin .cm
\evensidemargin .cm
\usepackage[numbers,sort&compress]{natbib}
\begin{document}

  \title{Third-order elastic coefficients and logarithmic strain\\ 
in constitutive and finite element modelling\\ of anisotropic
  elasticity} 
  \author{ {Pawe\l{} D\l{}u\.{z}ewski},{Marcin Ma\'{z}dziarz}, Piotr Tauzowski\\
    Institute of Fundamental Technological Research (IPPT PAN)\\ ul.\,Pawi\'{n}skiego 5$^{\rm B}$,
    02-106 Warszawa, email: pdluzew@ippt.pan.pl}
          
\maketitle
\begin{abstract}
  Third-order elastic coefficients (TOECs) have been measured experimentally and tabulated with pretty good accuracy since
  the middle of the previous century. In the classical acoustic measurement method the recalculation of instantaneous
  stiffness change onto TOECs is based on the use of Green strain. In recent calculations performed by means of atomistic
  and quantum methods many different strain measures are employed. In result, quite different sets of TOECs can be
  obtained for the same material. In this paper, it is shown how dramatically the coefficients obtained depend on the choice
  of strain measure. The known formulas for calculation of the second derivative of a tensor-valued function of tensor
  variable are corrected. The formulas are essential for the correct analytic calculation of the tangent
  stiffness matrix in finite element method, among others.

\end{abstract}


\section{Introduction}
Recently, the classical second-order elastic constants (SOECs) as well as the higher-order elastic coefficients are more and
more often predicted theoretically on the basis of quantum calculations
\cite{Cousins03,ZhaoW07,LopuszynskiM07,WangL09,SinghP11,WangW10,WangG13,JonesM14,HmielW16}. Over many years the elastic
coefficients were measured mainly by means of acoustic measurement methods. Thanks to such measurements the second-- and
third--order elastic coefficients have been tabulated {\em with pretty good accuracy} for the most of known crystal
structures. Despite so large database of TOECs collected over many years, still every year tens of new papers presenting
TOECs for known and novel materials are published in journals of solid physics.

It is worth mentioning that {\em anisotropic} hyperelastic models fitted to TOECs compose only a very small subset among
numerous constitutive models describing the anisotropic elastic behaviour of solids. In continuum mechanics a term {\em
  nonlinear elasticity} is often identified not with the quantitative modelling of elastic softening or hardening but with
a modelling in the limits of finite elastic deformations. Potentially, most of such obtained models can be fitted to the
experimentally determined TOECs. Nevertheless, it is easy to note that these two groups of papers develop rather
separately. The present paper is situated on the border of these two groups.

Constitutive models of elastic materials should have two fundamental properties:
\begin{itemize}
\item First of all, such models should hold the energy balance for an arbitrarily chosen elastic deformation
  loop. Unfortunately, many of elastic models ignore this fundamental rule. This concerns most of so-called Cauchy
  elastic materials for which the constitutive relationship is stated directly between the stress and strain measures, see
  \cite{Carroll09}. A similar problem does concern the linear elasticity. Despite that a linear theory is regarded as a
  theory which does not distinguish the Lagrangian and Eulerian configurations (small strain approach), the correct
  integration of elastic work done should take into account the change of elastic body shape. As a matter of fact, for each
  of deformation problem the linear theory of elasticity can be used to obtain two mutually different analytic solutions for
  displacement vector field derived alternatively from the Lagrangian or Eulerian configuration. For example, the most of
  analytic solutions applied in civil engineering satisfy the boundary conditions assumed in the Lagrangian
  configuration. Contrary to civil engineering, almost all of the analytic solutions available in the linear theory of
  dislocations satisfy the boundary conditions imposed in the Eulerian configuration, cf.\ e.g.\ Eqs (3-45) and (3-46) in
  \cite{HirthL82}. The elastic spring back of such a self-stressed structure leads to quite different analytic function for
  displacement vector derived from the Lagrangian configuration being then a single-- or a multi--connected region of a
  perfect lattice, cf.\ \cite{CholewinskiM14}.

\item A second but no less important issue is that the nonlinear elastic behaviour of constitutive models should be in
  agreement with the nonlinearity demonstrated by real materials. For example, the best-known {\em anisotropic} Hookean
  based on Green strain demonstrates a very strong change of instantaneous stiffness. Unfortunately, this elastic hardening
  runs in the quite opposite direction than those demonstrated by real materials. Namely, according to the experimental
  evidence, {\em the instantaneous stiffness of crystals increases strongly under compression and decreases under
    tension}. This phenomenon is responsible for many physical effects. For example, the experimentally observed sensitivity
  of elastic stiffness on the applied stress has been used in commercial devices for detection of the residual stresses by
  means of acoustic waves \cite{EgleB69,DeputatS92,Schramm99,Chipanga10}. Another domain of research in which TOECs play an
  important role is optoelectronics. Namely, a GPa stress assisting the lattice mismatch between monolayers in semiconductor
  devices often results in a strong piezoelectric effect. This effect shifts the band gap in quantum structures and changes
  their optical properties, {\it cf}.\,\cite{FrogleyD00,Lepkowski08}.  Recently, the effect of the higher order elastic and
  piezoelectric coefficients on the optical properties of quantum dots is calculated numerically, see
  e.g.\,\cite{Jurczak16,JurczakD18}. Another example concerns the nonlinear elastic effect induced by dislocations. Due to
  the asymmetry in the stress/strain response for the extension and compression, the nonlinear theory of crystal defects
  predicts the volume expansion induced by edge dislocations and other extended defects, see
  \cite{HorodonA61,HirthL82,Spaepen00}. TOECs are used also for prediction of elastoplastic behaviour of materials
  undergoing megabar pressures up to 300\,GPa \cite{Nielsen86,Clayton14,FengL16}. In geophysics, an isothermal Eulerian
  Birch--Murnaghan equation of state is used despite of the recognized shortcomings for large compressive strains
  \cite{Murnaghan37,Birch47}. Poirier et al.\,\cite{Poirier98} have shown that the analogical state equation for strain
  energy constructed from logarithmic (Hencky) strain is more stable than the traditional Eulerian Birch--Murnaghan equation
  based on the use of Green strain.
\end{itemize}

The awareness of a strong dependency of TOECs on the choice of strain measure, could be the reason why
\citeauthor{ThurstonS66} already in 1966 named TOECs not {\em constants} but only {\em coefficients}. Since that time, TOECs
have been successfully determined by measuring the dependency of elastic wave speed on the stress applied to crystal
structures \cite{Brugger64,ThurstonB64,HikiG66,Walker85,HankeyS70,Wallace70,JohalD06,SinghS08,DestradeO10,LangG11}. Despite
the fact that the measurement of higher order elastic coefficients of this kind has a long tradition, it appears
occasionally that some formulas used through so many years and settled down as a classic are not quite correct. For example
recently, the formulas for the extraction of pressure coefficients derived by McSkimin and Andreatch in 1972
\cite{McSkiminA72} were corrected by Winey et al. \cite{WineyH16}. In the present paper a similar situation is shown with the
formulas for the second order derivative of a tensor-valued function of a symmetric second--order tensor.

In the next section the deformation and strain measures used in the linear and finite deformation theories of elasticity are
discussed in brief. Special attention is focused on the strict mathematical relationships between so called linear strain
measures and finite strain measures related to mutually different reference configurations. In the mentioned section the
known formulas for calculation of the second derivative of a tensor-valued function of tensor variable are
corrected. Section 3 is devoted to anisotropic hyperelastic models in which TOECs are used to fit the behaviour of
constitutive models to the anisotropic elastic behaviour of real materials. The effect of strain measure on the resultant
elastic hardening of Hookian models is discussed in terms of experimentally determined elastic hardening data. This
hardening is quantified by means of TOECs fixed experimentally in relation to an arbitrary chosen strain measure. With
respect to the revision of the formulas mentioned above, in Section 4 the revised formulas are used for calculation of
tangent stiffness matrix in finite element method.

\section{Deformation measures} 
In continuum mechanics two main approaches to the modelling of elastic body are applied. In the linear approach, the body
motion is described by means of a displacement vector function that depends either on the reference position or the spatial
position of a material particle. Contrary to that, in nonlinear theories the current position $\mathbf x$ of the material
particle is assumed to be a function of the reference position, ${\mathbf x}={\mathbf x}({\mathbf X})$. In such
a case, $\mathbf x$ and $\mathbf X$ are treated as points given by means of curvilinear coordinates on a differential
manifold. Neglecting some special cases like the continuum theory of discrete dislocations, the mentioned mapping is assumed
to be regular and reversible which means that in most cases an inverse function ${\mathbf X}={\mathbf X}({\mathbf x})$
is assumed to be regular and invertible. The last approach based on the use of manifolds is found to be more general with
respect to the possibility of considering the deformation of an elastic body in a curvilinear space where a displacement
vector field of the same order as of the order of the manifold may not exist. In the present paper, we consider finite
deformation of elastic body by means of curvilinear coordinates embedded in a 3D Euclidean space. In more general case, the
analytic relations presented here can be easy translated to an arbitrary 3D Riemannian manifold on which a parallel
connection is additionally determined.

Taking into account that ${\mathbf X}= {\mathbf x} - {\mathbf u}$, the {\em mutually reversible} relationship can be
rewritten in the form
\begin{align}
{\mathbf F}= \big({\mathbf 1}&-\nabla{\mathbf u} \big)^{-1}= {\mathbf 1}+\widehat{\nabla}{\mathbf u}\,,&\nabla{\mathbf u} = {\mathbf 1} &- {\mathbf F}^{-1}\,,\label{Fx-1}&
  \widehat{\nabla}{\mathbf u} = {\mathbf F}&-{\mathbf 1}\,,
\end{align}
where ${\mathbf F}=\fracp{\mathbf x}{\mathbf X}$, \ ${\mathbf F}^{-1}\!=\fracp{\mathbf X}{\mathbf x}$; and $\mathbf 1$ denotes the second order unit
tensor.

According to the theorem of polar decomposition, the deformation gradient can be decomposed into the orthogonal tensor of
rotation, $\mathbf R$, and the right or left stretch tensors, respectively
\begin{align}
\mathbf F = \mathbf R\mathbf U = {\mathbf V\mathbf R}. \label{Fechpl}
\end{align}
Contrary to that, in the linear theory, the displacement gradient is decomposed into an antisymmetric rotation tensor,
${\mathbf w}$, and symmetric strain tensor, ${\boldsymbol\varepsilon}$, referred alternatively to the Eulerian (spatial) or
Lagrangian (reference) configurations,
\begin{align}
\nabla {\mathbf u} &= {\mathbf w} + {\boldsymbol\varepsilon}\,, 
& \widehat{\nabla} {\mathbf u} &= \widehat{\mathbf w} + \widehat{\boldsymbol\varepsilon}\,, \label{uew}
\end{align} 
where
\begin{align} 
{\mathbf w}& \stackrel{df}{=}\textstyle\frac{1}{2}\big(\partial_{\mathbf x}{\mathbf u} - \partial^T_{\mathbf x}{\mathbf u}\big)\,,&
{\boldsymbol\varepsilon}& \stackrel{df}{=}\textstyle\frac{1}{2}\big(\partial_{\mathbf x}{\mathbf u} + \partial^T_{\mathbf x}{\mathbf u}\big),\label{we}\\
\widehat{\mathbf w}&\stackrel{df}{=}\textstyle\frac12\big(\partial_{\mathbf X}{\mathbf u} -\partial^T_{\mathbf X}{\mathbf u}\big),&
\widehat{\boldsymbol\varepsilon}&\stackrel{df}{=}\textstyle\frac12\big(\partial_{\mathbf X}{\mathbf u} +\partial^T_{\mathbf X}{\mathbf u}\big),
\end{align}
the displacement vector function is treated then as a composite vector function
${\mathbf u}={\mathbf f}\big({\mathbf x}({\mathbf X})\big)$ or ${\mathbf u}=\widehat{\mathbf f}({\mathbf X})$,
respectively. A linear theory of elasticity is often regarded as a theory that by definition ignores any difference between
the reference and spatial configuration of elastic body. As a matter of fact, in practical uses, the analytic solutions of
boundary-value problems obtained by means of the linear theory can be divided into two separated groups that satisfy the
imposed symmetry (loading, kinematic constrains, boundary conditions) in the reference or spatial configurations,
alternatively. For example, in the linear theory of dislocations, almost all the known analytic solutions for stress-stain
fields have been obtained for the Eulerian configuration. In such a case, in order to find the reference position of a
material particle in the Lagrangian configuration the displacement vector field must be pulled back from the Eulerian
configuration, {\em cf.}  the analytic solutions for dislocations in \cite{HirthL82}. Recently, the lack of analytic
solutions for the displacement vector field related directly to the Lagrangian configuration poses some problem in the use
of analytic formulas for the preprocessing of atomistic models of dislocation networks in noncrystalline heterostructures,
cf.\ \cite{YoungK07,CholewinskiM14}.

{\bf Example} {\small\em To illustrate the problem of inaccuracies in the determination of strain with the use of linear
  strain measures, let us consider two different gradients of displacement vector
\begin{align}
[\nabla {\mathbf u}_1] &= \Bigg[ \begin{smallmatrix} 
\frac 1 2        &\frac {\sqrt{3}} 2& 0\\
-\frac{\sqrt{3}} 2&    \frac 1 2      & 0\\ 
 0               &   0               & 1 
\end{smallmatrix}\Bigg]\,,&
\Bigg[ \begin{smallmatrix} 
-\frac 1 2  & \frac{\sqrt{3}} 2 & 0\\
-\frac{\sqrt{3}} 2 & -\frac 1 2   & 0\\
 0                &   0         & 1 
\end{smallmatrix}\Bigg]&= [\nabla{\mathbf u}_2] .\label{u12}
\end{align} 
From the view point of linear theory the additive decomposition into antisymmetric and symmetric parts gives 
\begin{align}
[{\boldsymbol\varepsilon}_1] &=\Bigg[ \begin{smallmatrix} 
-\frac 1 2 & 0    & 0 \\
0   & -\frac 1 2  & 0 \\
0&0& 0 
\end{smallmatrix}\Bigg],&
{\mathbf w}_1&=\Bigg[ \begin{smallmatrix} 
0  & \frac{\sqrt{3}} 2 & 0 \\
-\frac{\sqrt{3}} 2& 0 & 0  \\
 0            &0 & 0 
\end{smallmatrix}\Bigg]={\mathbf w}_2\,,&
 \Bigg[ \begin{smallmatrix} 
\frac 3 2 & 0    & 0 \\
0   & \frac 3 2  & 0 \\
0&0& 0 
\end{smallmatrix}\Bigg] &=[{\boldsymbol\varepsilon}_2]\,.
\end{align} 
From the viewpoint of finite deformation theory the substitution of \eqref{u12} into (\ref{Fx-1}b) gives the
following matrix representations of deformation gradient
\begin{align}
[{\mathbf F}_1] &= \Bigg[ \begin{smallmatrix} 
\frac 1 2        &-\frac {\sqrt{3}} 2& 0\\
\frac{\sqrt{3}} 2&    \frac 1 2      & 0\\ 
 0               &   0               & 1 
\end{smallmatrix}\Bigg]\,,&
\Bigg[ \begin{smallmatrix} 
-\frac 1 2  & -\frac{\sqrt{3}} 2 & 0\\
\frac{\sqrt{3}} 2 & -\frac 1 2   & 0\\
 0                &   0         & 1 
\end{smallmatrix}\Bigg]& = [{\mathbf F}_2] .\label{F12}
\end{align} 
One can check that these two mutually different deformation gradients are nothing more than two rigid rotations around $z$
axis through the angles $\frac \pi 6$ and $\frac {5\pi} 6$, respectively. For a rigid body rotation, the strain tensors
should vanish. Thus, the non-zero components of linear strains $[{\boldsymbol\varepsilon}_1]$ and
$[{\boldsymbol\varepsilon}_2]$ represent only errors generated by the linear theory. As a matter of fact, the anti symmetric
tensors ${\mathbf w}_1={\mathbf w}_2$ do not distinguish these two quite different rotations.}

One of important consequence of the inaccurate determination of strain in linear theories is the incorrect determination of
strain energy --- independently of whether the elastic strain is very small or not. We do not discuss here which level of
linear elastic strain is a sufficiently small to neglect the perpetual motion of such elastic models on deformation
loops. The important here is that the mathematical dependency of linear strain measures on rigid rotation makes impossible
the measurement of many material coefficients with the use of these strain measures. This concerns TOECs as well as many
other coefficients like the second and higher order piezoelectric and magnetostriction coefficients.

\subsection{Logarithmic strain measure based on polar decomposition}
According to the spectral decomposition theorem the right stretch tensor can be decomposed into three real and positive
eigenvalues, $u_1,u_2,u_3$, and three unit eigenvectors ${\mathbf u}_1,{\mathbf u}_2,{\mathbf u}_3$, which can be used for
the following decomposition of the stretch tensor
\begin{align}
{\mathbf U}& = \sum_{{\sf n}=1}^3u_{\sf n} \, {\mathbf u}_{\sf n}\otimes {\mathbf u}_{\sf n}, 
\end{align}
where $\otimes$ denotes the dyadic product.
\paragraph{\bf Definition}
\label{defin} {\em By a Lagrangian strain measure we mean a tensor-valued isotropic function of the right stretch tensor,
\begin{align}
\widehat{{\boldsymbol\varepsilon}}&\stackrel{df}{=}\sum_{{\sf n}=1}^3f(u_{\sf n}) \ {\mathbf u}_{\sf n}\otimes {\mathbf u}_{\sf n}, \label{e(u)}
\end{align}
where $f(\cdot)$ is an arbitrarily chosen, sufficiently smooth and monotonically increasing function $f(x):{\mathbb
  R}^{+}\ni x \rightarrow f\in {\mathbb R}$ which satisfies the conditions: $f(x)|_{x=1} ={0}$ and $\bigl. \frac{df(x)}{dx}
\bigr|_{x=1}=1$.\\} In this definition ${\mathbb R}$ and ${\mathbb R}^+$ denotes the sets of real and positive real numbers,
respectively. This definition includes among others the well-known Seth-Hill family of strain measures defined as
\begin{align}
\widehat{\boldsymbol\varepsilon} &\stackrel{df}{=} \begin{cases}
\frac{1}{m} ({\mathbf U}^m - {\mathbf 1}) & \text{for}\quad m\ne 0,\\
\ln{\mathbf U} & \text{for}\quad m=0, \end{cases}   \label{eUn}
\end{align}
For $m = 2, 1, 0$ and -2 the Green, Biot, logarithmic (Hencky) and Almansi strain measures are obtained, respectively.
For a uniaxial stretch test, these family of strain measures was gathered into a common formula by \citet{Seth62} in
\citeyear{Seth62}. In \citeyear{Hill70} \citet{Hill70} first applied a definition of an analytic function of a matrix for
generalization the Seth's definition to a 3D case. As a matter of fact, functions of matrices have been studied for as long
as matrix algebra itself \citep{Gantmacher54,Higham08}. Indeed, in his seminal {\em A Memoir on the Theory of Matrices}
({\bf 18}58), Cayley investigated the square root of a matrix, and it was not long before definitions of
${\mathbf f}({\mathbf U})$ for general $\mathbf f$ were proposed by Sylvester and others, see \citet{Higham08}.  Seth-Hill
family of strains by no means fulfils all possible measures of \eqref{e(u)}, cf.\,\cite{CurnierZ06}.

As shown by \citet{Hill78}, the material derivative of (\ref{e(u)}) reads
\begin{align}
\dot{\widehat{\boldsymbol\varepsilon}} &= \ \ \,\widehat{\pmb{\!\!\mathcal A\ \ }}\!\!\! :({\mathbf R}^{T}{\mathbf d}\, {\mathbf R}), &\text{where} 
&& 
{\mathbf d} &\stackrel{df}{=} \frac 1 2 \big[  \dot{\mathbf F}{\mathbf F}^{-1} + (\dot{\mathbf F}{\mathbf F}^{-1})^T \big] .\label{eeq3} 
\end{align}
In the eigenvector basis $\{ {\mathbf u}_{\sf K}\}$ the fourth--order proper--symmetric tensor $\ \widehat{\pmb{\!\!\mathcal A\ \ }}\!\!\!$
is represented by the following non-vanishing components \cite{Dlu00}
\begin{equation}
  {\mathcal{\widehat A}}_{\sf ijij} = {\widehat{\mathcal A}}_{\sf ijij} =
\begin{cases} 
 u_{\sf i} f'(u_{\sf j}) & \text{for ${\sf i} = {\sf j}$,} \\
\frac 1 2  u_{\sf i} f'(u_{\sf j}) & \text{for $ u_{\sf i} =  u_{\sf j}$ and ${\sf i} \ne {\sf j}$,} \\
\frac{u_{\sf i}u_{\sf j}[f(u_{\sf i})-f(u_{\sf j})]}{u^{2}_{\sf i}-u^{2}_{\sf j}} 
 & \text{for $ u_{\sf i}\ne  u_{\sf j}$.}
\end{cases}\label{dfT1} 
\end{equation}
sans serif fonts are reserved here for components rewritten in a special vector basis composed of the stretch eigenvectors
{\it cf.}\ \eqref{dfT1}. For the Hencky and Green strain measures this function reads, respectively
\begin{align}
\widehat{\mathcal A}_{\sf ijij} &= 
\begin{cases}  
1 & \text{for ${\sf i}={\sf j}$,} \\ 
\frac 1 2 & \text{for $ \widehat{\varepsilon}_{\sf i} = \widehat{\varepsilon}_{\sf j}$ and ${\sf i}\ne{\sf j}$,} \\ 
\frac{\widehat{\varepsilon}_{\sf i}-\widehat{\varepsilon}_{\sf j}}{2\sinh{(\widehat{\varepsilon}_{\sf i}-\widehat{\varepsilon}_{\sf j})}}  
 & \text{for $ \widehat{\varepsilon}_{\sf i}\ne \widehat{\varepsilon}_{\sf j}$,} 
\end{cases}  
&
\widehat{\mathcal A}_{\sf ijij} &=
\begin{cases}  
u_{\sf i}^2 & \text{for ${\sf i}={\sf j}$ ,} \\ 
\frac 1 2 u_{\sf j} u_{\sf i} & \text{for ${\sf i}\ne {\sf j}$ .}
\end{cases}  \label{deft} 
\end{align} 
In Voigt notation the symmetric stress and strain tensors are treated as six-dimensional vectors in which the shear strain
components take double values. Alternatively, in the Mandel notation the shear components both of the 6-dimensional stress
and strain vectors are multiplied equally by $\sqrt{2}$. It is worth mentioning that representations of symmetric
second-order tensors satisfy all postulates of the Euclidean six dimensional space. Thus, the two notations mentioned above
are identified here with two representations of 6 dimensional vectors in two mutually different vector bases of a 6D
Euclidean space, respectively. In order to distinguish components of vectors and tensors decomposed in
${\mathsf E}^3\otimes{\mathsf E}^3$ and ${\mathsf E}^6$, the elements of vector bases in ${\mathsf E}^6$ will be denoted
here by Gothic characters, e.g.\, ${\pmb{\mathfrak e}}_{\ae}$ vs ${\mathbf e}_{\textrm a}$ and ${\mathbf e}_{\textrm e}$.

In the case of an orthonormal coordinate system in ${\mathsf E}^3$ the scalar product of the stress and strain tensors used
to determine the strain energy can be rewritten in the following form
\begin{align}
\boldsymbol\sigma\!:\!\boldsymbol\varepsilon
&= \sigma_{11}\varepsilon_{11} + \sigma_{22}\varepsilon_{22} + \sigma_{33}\varepsilon_{33} 
+2 \sigma_{23}\varepsilon_{23} + 2 \sigma_{31}\varepsilon_{31} + 2 \sigma_{12}\varepsilon_{12}\,.\label{psi6}
\end{align}

Representations of the tensors in Voigt notation will be identified here with the covariant and contravariant components of
the following six--dimensional vectors
\begin{subequations}
\begin{align}
[\varepsilon^{i\mspace{-4.5mu}j}]^T &= [\varepsilon^1,\varepsilon^2,\varepsilon^3,\varepsilon^4,\varepsilon^5,\varepsilon^6]^T
\equiv [\varepsilon_{11},\varepsilon_{22},\varepsilon_{33},2\varepsilon_{23},2\varepsilon_{13},2\varepsilon_{21}]^T,\\
[\sigma_{i\mspace{-4.5mu}j}]  &= [\sigma_1,\sigma_2,\sigma_3,\sigma_4,\sigma_5,\sigma_6]\equiv[\sigma_{11},\sigma_{22},\sigma_{33},\sigma_{23},\sigma_{13},\sigma_{21}]\,.
\end{align}
\end{subequations}
In such a case, the right hand of \eqref{psi6} can be treated as a scalar product of vector components rewritten in
the Voigt vector basis for which the co-- and contravariant representations of the unit tensor read
\begin{align}
[{\mathfrak g}_{{i\mspace{-4.5mu}j}\ae}]&=\begin{bmatrix} 
1&\cdot&\cdot&\cdot&\cdot&\cdot\\
\cdot&1&\cdot&\cdot&\cdot&\cdot\\
\cdot&\cdot&1&\cdot&\cdot&\cdot\\
\cdot&\cdot&\cdot&\frac 1 2 &\cdot&\cdot\\
\cdot&\cdot&\cdot&\cdot&\frac 1 2 &\cdot\\
\cdot&\cdot&\cdot&\cdot&\cdot&\frac 1 2 
\end{bmatrix}\,,&
[{\mathfrak g}^{{i\mspace{-4.5mu}j}\ae}]&=\begin{bmatrix} 
1&\cdot&\cdot&\cdot&\cdot&\cdot\\
\cdot&1&\cdot&\cdot&\cdot&\cdot\\
\cdot&\cdot&1&\cdot&\cdot&\cdot\\
\cdot&\cdot&\cdot&2&\cdot&\cdot\\
\cdot&\cdot&\cdot&\cdot&2&\cdot\\
\cdot&\cdot&\cdot&\cdot&\cdot&2 \end{bmatrix}\,, \label{gotg}
\end{align}
where $i\mspace{-4.5mu}j$ denotes the Voigt type index corresponding to a pair of subscripts $(i,j)$ used for tensors in
${\mathsf E}^3\otimes{\mathsf E}^3$. In our notation, the pair of indices and its six dimensional counterpart in
${\mathsf E}^6$ will be denoted by analogical symbols, e.g.\ $ii$ and $\textrm a\textrm e$ will be replaced their ligatures
$\ddot u$, $\ae$, or simply by diacritical symbols obtained by overlapping the original ones, $i\mspace{-1.1mu} i$ and
$\textrm a\mspace{-3.5mu}\textrm e$. By analogy to classical force and displacement components, the components $\sigma_{ij}$
and $\varepsilon_{ij}$ will be identified here respectively with co-- and contravariant representations of six dimensional
stress and strain vectors, $\sigma_{i\mspace{-4.5mu}j}$ and $\varepsilon^{i\mspace{-4.5mu}j}$, respectively. For example,
the volume density of strain energy can be rewritten in the form
\begin{gather}
  \psi_V = \frac 1 2 \, {\boldsymbol\sigma}\!:\!{\boldsymbol\varepsilon} 
= \frac 1 2 \, \sigma_{i\mspace{-4.5mu}j}\,\varepsilon^{i\mspace{-4.5mu}j} = \frac 1 2\, \sigma_{ij}\,\varepsilon^{ij}.
\end{gather}

\subsection{Derivatives of tensor-valued functions}
In Voigt notation the contravariant, covariant and mixed representations of the first derivative of a tensor-valued function
of a symmetric second-order tensor, ${\mathbf f}({\mathbf U})$, can be rewritten in the following form
\begin{align} 
\fracp{f^{i\mspace{-4.5mu}j}}{U_{i\mspace{-4.5mu}j}} &=
    \begin{cases}  
      \ \, f'(\lambda_{\sf i})  & \text{for}\quad {\sf i} = {\sf j}, \\
      \ \, 2\, f'(\lambda_{\sf i})  & \text{for}\begin{cases}\lambda_{\sf i} = \lambda_{\sf j}\\i\ne j,\end{cases}\\
      2 \frac{f(\lambda_{\sf i})-f(\lambda_{\sf j})}{\lambda_{\sf i}-\lambda_{\sf j}}&\text{for}\quad\lambda_{\sf i}\ne\lambda_{\sf j},
    \end{cases} 
    \label{dfT}\\
\fracp{f_{i\mspace{-4.5mu}j}}{U^{i\mspace{-4.5mu}j}} &=
    \begin{cases}  
      \ \, f'(\lambda_{\sf i})  & \text{for}\quad {\sf i} = {\sf j}, \\
      \ \, \frac12 f'(\lambda_{\sf i})  & \text{for}\begin{cases}\lambda_{\sf i} = \lambda_{\sf j}\\i\ne j,\end{cases}\\
      \frac12 \frac{f(\lambda_{\sf i})-f(\lambda_{\sf j})}{\lambda_{\sf i}-\lambda_{\sf j}}&\text{for}\quad\lambda_{\sf i}\ne\lambda_{\sf j}.
    \end{cases} \\
\fracp{f^{i\mspace{-4.5mu}j}}{U^{i\mspace{-4.5mu}j}} = \fracp{f_{i\mspace{-4.5mu}j}}{U_{i\mspace{-4.5mu}j}} &=
    \begin{cases}  
      \ \, f'(\lambda_{\sf i})  & \text{for}\quad\lambda_{\sf i} = \lambda_{\sf j}, \\
     \frac{f(\lambda_{\sf i})-f(\lambda_{\sf j})}{\lambda_{\sf i}-\lambda_{\sf j}} & \text{for}\quad\lambda_{\sf i}\ne \lambda_{\sf j},
    \end{cases} \label{d1fTm}
\end{align} 
The mentioned representations satisfy the following transformation rule
$\fracp{f^{\ae}}{U^{d\!\!z}} ={\mathfrak g}^{\ae i\mspace{-4.5mu}j} \fracp{f_{i\mspace{-4.5mu}j}}{U_{\mathit{fl}}}
{\mathfrak g}_{\mathit{fl} d\!\!z} = {\mathfrak g}^{\ae i\mspace{-4.5mu}j} \fracp{f_{i\mspace{-4.5mu}j}}{U^{\mathit{fl}}}
{\mathfrak g}{^{\mathit{fl}}}\!\!_{d\!\!z} = {\mathfrak g}{^{\ae}}\!_{i\mspace{-4.5mu}j}
\fracp{f^{i\mspace{-4.5mu}j}}{U_{\mathit{fl}}} {\mathfrak g}_{\mathit{fl} d\!\!z}$. A 6x6 dimensional unit tensor
$\pmb{\mathfrak g}$ can be spanned on two different vector bases. Then, the shifters
${\mathfrak g}{^{\ae'}}\!_{i\mspace{-4.5mu}j}$ give possibility to transform the tensor components from one vector basis to
another, e.g.\ from the Voigt vector basis to the orthonormal one in $\sf E^6$.

Analytic formulas for $\fracp{\mathbf f^2}{\mathbf U^2}$ take a more complex form than those for
$\fracp{\mathbf f}{\mathbf U}$. The first derivation of such analytic formulas (not quite correct) is due to
\citet{BowenW70}, see (3.50-3.61) therein. In 1971 \citeauthor{ChadwickO71} presented the results of an independent
calculation in which the main errors of \citet{BowenW70} were corrected. As a matter of fact, they repeated some errors
after \citeauthor{BowenW70}. In result, it seems that in none of papers published as of yet the mentioned equations have
been corrected. In \citeyear{Norris08}, \citet{Norris08} presented in a smart form a general mathematical scheme for
calculation of an n$^{th}$--order derivative, see\ (3.16) therein. The present authors, not being quite convinced which form
of the formulas for the second derivative is correct, have performed a series of the validation tests. Due to the
non--orthonormality of the Voigt vector basis $\{{\frak e}_{\ae}\}$, the first test was performed for the mixed
representation by means of the standard finite difference scheme adopted here to ${\sf E}^6$,
\begin{subequations} 
\begin{align}
  \frac{\partial^2f_{i\mspace{-4.5mu}j}}{\partial U_{\ae}\partial U_{\ae}} &=\frac{f_{i\mspace{-4.5mu}j}({\mathbf U}+{\boldsymbol\Delta}_{\ae})
    -2f_{i\mspace{-4.5mu}j}({\mathbf U})+f_{i\mspace{-4.5mu}j}({\mathbf U}-{\boldsymbol\Delta}_{\ae})} {\Delta_{\ae} \Delta_{\ae}},\\
  \frac{\partial^2f_{i\mspace{-4.5mu}j}}{\partial U_{\ae}\partial U_{\mathit{fl}}}&=\frac{f_{i\mspace{-4.5mu}j}({\bf
      U}+{\boldsymbol\Delta}_{\ae}+{\boldsymbol\Delta}_{\mathit{fl}}) -f_{i\mspace{-4.5mu}j}({\bf
      U}-{\boldsymbol\Delta}_{\ae}+{\boldsymbol\Delta}_{\mathit{fl}}) -f_{i\mspace{-4.5mu}j}({\bf
      U}+{\boldsymbol\Delta}_{\ae}-{\boldsymbol\Delta}_{\mathit{fl}}) +f_{i\mspace{-4.5mu}j}({\bf
      U}-{\boldsymbol\Delta}_{\ae}-{\boldsymbol\Delta}_{\mathit{fl}})}{4\,\Delta_{\ae}
    \Delta_{\mathit{fl}}} \end{align} \end{subequations} where $[{\boldsymbol\Delta}_{\ae}]^T = [U_1,\ldots,\,
U_{\ae}+\Delta_{\ae}\ ,\ldots, U_6]^T$. After this test had passed, the following correct contravariant representation 
\eqref{d2fTa} was fixed by using the chain rule $\frac{\partial^2f^{\ddot{u}}} {\partial U^{\ae}\partial U^{\mathit{fl}}} = {\frak
  g}^{\ddot{u}\,i\mspace{-4.5mu}j} \frac{\partial^2f_{i\mspace{-4.5mu}j}} {\partial U^{\ae}\partial U^{\mathit{fl}}}$, 
\begin{gather} \fbox{$\displaystyle
\begin{aligned}
\fracp{^2f^{i\mspace{-4.5mu}j}}{U_{j\mspace{-3.0mu}k}\partial U_{k\mspace{-2.0mu}i}} &= \begin{cases}  
f''(\lambda_{\sf i})  & \text{for}\quad \lambda_{\sf i}= \lambda_{\sf j}= \lambda_{\sf k} \,,\\
2\frac{f'(\lambda_{\sf j}) - \frac{f(\lambda_{\sf i})-f(\lambda_{\sf j})}{\lambda_{\sf i}-\lambda_{\sf j}}}{(\lambda_{\sf i}-\lambda_{\sf j})}
  & \text{for}\quad \lambda_{\sf i}\ne \lambda_{\sf j} = \lambda_{\sf k} \,,\\
-2\frac{f(\lambda_{\sf i}) ( \lambda_{\sf j}-\lambda_{\sf k}) + f(\lambda_{\sf j}) ( \lambda_{\sf k}-\lambda_{\sf i}) + f(\lambda_{\sf k}) ( \lambda_{\sf i}-\lambda_{\sf j} )}
     {(\lambda_{\sf i}-\lambda_{\sf j})( \lambda_{\sf j}-\lambda_{\sf k})( \lambda_{\sf k}-\lambda_{\sf i})}
  & \text{for}\quad \lambda_{\sf i}\ne \lambda_{\sf j}\ne \lambda_{\sf k}\ne  \lambda_{\sf i}\,, 
\end{cases} 
\end{aligned}$\!\!\!\!} 
\label{d2fTa}
\end{gather}
where the subscripts $\sf i,j,k$ run over all possible values: 1,2,3. In order to compare our results with those presented
in the previous papers it is convenient to rewrite
${\cal L}^{(2)}_{ijklmn} =\fracp{^2f_{ij}}{U^{kl}\,\partial U^{mn}}$ in the following form
\begin{subequations}\label{d2fTbd}
  \begin{align} 
    {\cal L}^{(2)}_{iiiiii} &=   f''(\lambda_{\sf i})\,,  & \label{d2fTb}\\
    {\cal L}^{(2)}_{iiijij} &= \begin{cases}  
      \frac{\mathbf 1}{\mathbf 4} f''(\lambda_{\sf i})  & \text{for}\quad \lambda_{\sf i}= \lambda_{\sf j} \,,\\
      \frac{\mathbf 1}{\mathbf 2}\frac{\frac{f(\lambda_{\sf i})-f(\lambda_{\sf j})}{\lambda_{\sf i}-\lambda_{\sf j}} - f'(\lambda_{\sf i})}{(\lambda_{\sf i}-\lambda_{\sf j})}
      & \text{for}\quad \lambda_{\sf i}\ne \lambda_{\sf j}\,,
    \end{cases} 
        \label{d2fTc}\\
    {\cal L}^{(2)}_{ijjkki} &= \begin{cases}  
      \frac18f''(\lambda_{\sf i})  & \text{for}\quad \lambda_{\sf i}= \lambda_{\sf j}= \lambda_{\sf k} \,,\\
      \frac14\frac{\frac{f(\lambda_{\sf j})-f(\lambda_{\sf i})}{\lambda_{\sf j}-\lambda_{\sf i}} - f'(\lambda_{\sf j})}{(\lambda_{\sf j}-\lambda_{\sf i})}
      & \text{for}\quad \lambda_{\sf i}\ne \lambda_{\sf j} = \lambda_{\sf k} \,,\\
      -\frac14\frac{f(\lambda_{\sf i}) ( \lambda_{\sf j}-\lambda_{\sf k}) + f(\lambda_{\sf j}) ( \lambda_{\sf k}-\lambda_{\sf i}) + f(\lambda_{\sf k}) ( \lambda_{\sf i}-\lambda_{\sf j} )}
      {(\lambda_{\sf i}-\lambda_{\sf j})( \lambda_{\sf j}-\lambda_{\sf k})( \lambda_{\sf k}-\lambda_{\sf i})}
      & \text{for}\quad \lambda_{\sf i}\ne \lambda_{\sf j}\ne \lambda_{\sf k}\ne  \lambda_{\sf i}\,, 
    \end{cases} 
        \label{d2fTd}
  \end{align}
\end{subequations}
where only $\sf i\ne j\ne k$ take possible values: 1,2,3. The terms corrected here have been rewritten in bold,
cf.\,(3.5.33--34) in \cite{Ogden84}. The question then arises how it was possible that so many years a formula so important
for calculation of the second derivatives of a tensor-valued functions have not been corrected. Recently, most of the
complex analytic differentiation is computed by means of programs for symbolic calculations. Such programs are very useful
under one condition, namely, that somebody earlier has programmed a given analytic operation. Otherwise, there is rather a
small chance that a given error will be detected by means of elemental operations already programmed in.

\section{Anisotropic hyperelasticity and TOECs in terms of logarithmic strain}
In the case of quasistatic elastic deformation of an anisotropic body, the local form of the energy balance law can be
rewritten in the following way
\begin{equation}
-\rho \dot{\psi} + {\boldsymbol\sigma}:{\mathbf d}=0 ,\label{eeq}
\end{equation}
where $\rho,\dot{\psi}$ and $\boldsymbol\sigma$ respectively denote the mass density, material derivative of strain energy,
and the Cauchy stress. Suppose that the specific strain energy depends on the Lagrangian strain tensor
\begin{equation}
\psi=\psi(\widehat{\boldsymbol\varepsilon}) .\label{eeq0}
\end{equation}
Substitution of \eqref{eeq0} and \eqref{eeq3} into \eqref{eeq} gives
\begin{equation}
-\frac{\rho}{\widehat{\rho}}\  \Big( \widehat{\rho}\fracp{\displaystyle \psi}{\widehat{\boldsymbol\varepsilon}}\Big):
\ \,\widehat{\pmb{\!\!\mathcal A\ \ }}\!\!\! : ({\mathbf R}^{T}{\mathbf d}\, {\mathbf R}) + {\boldsymbol\sigma}:{\mathbf d}=0 ,\label{ee2}
\end{equation}
where $\widehat{\rho}=\rho \det{\mathbf F}$. To balance the energy for an arbitrarily chosen $\mathbf d$, the
Cauchy stress has to be governed by the following transformation rule
\begin{equation}\label{sra}
{\boldsymbol\sigma}={\mathbf R}\big(\ \,\widehat{\pmb{\!\!\mathcal A\ \ }}\!\!\!:\widehat{\boldsymbol\sigma}\big){\mathbf R}^{T} \det{\mathbf F}^{-1}, 
\end{equation}
where the stress conjugate to the given Lagrangian strain reads
\begin{align}
\widehat{\boldsymbol\sigma} &\stackrel{df} = \widehat{\rho}\, \fracp{\psi}{\widehat{\boldsymbol\varepsilon}}\,.\label{ws}
\end{align}

Constitutive equations for anisotropic Hookean model can be derived by assuming the following strain energy function
\begin{equation}
\widehat{\rho}\, \psi(\widehat{\boldsymbol\varepsilon})= 
\frac{1}{2}\, \widehat{\boldsymbol\varepsilon}:\widehat{\mathbf c}:\widehat{\boldsymbol\varepsilon} \label{psi1},
\end{equation}
where, by definition, $\widehat{\mathbf c}=\text{const.}$ The substitution subsequently into \eqref{ws} and \eqref{sra} gives
\begin{align}
\widehat{\boldsymbol\sigma}=\widehat{\mathbf c}:\widehat{\boldsymbol\varepsilon}&& 
\text{and}&&
{\boldsymbol\sigma}={\mathbf c}:{\boldsymbol\varepsilon}, \label{ca}  
\end{align}
where the Eulerian strain and stiffness tensors are defined by means of the following transformation rules
\begin{align}
{\boldsymbol\varepsilon}&\stackrel{df}{=} {\pmb{\mathcal{A}}}^{-1}\!:\widehat{\boldsymbol\varepsilon},\label{e=A-Ta}&
{\mathbf c}  &\stackrel{df}{=} {\pmb{\mathcal{A}}}^T\!:\widehat{\mathbf c}:{\pmb{\mathcal{A}}}  \det{\mathbf F}^{-1},
\end{align}
where ${\mathcal A}{_{IJ}}^{ij} \stackrel{df}{=} {R^{i}}_{K} {R^{j}}_{L}  \widehat{\mathcal A}{_{IJ}}^{KL}$\ \  and \ \ $\stackrel{-1}{\mathcal A}\!{_{ij}}^{IJ}= {R_{i}}^{K}{R_{j}}^{L}\!\stackrel{-1}{\widehat{\mathcal A}}\!{^{IJ}}\!_{KL}$.
According to this transformation rule, Almansi strain in the Eulerian
configuration is obtained as the Eulerian counterpart of Green strain, cf.\,Table \ref{Tb}. In the literature, different
notations and names are used. In the present paper, the tensors called the Eulerian and Lagrangian Almansi strains differ
each other only the rigid rotation.
\begin{table}
\begin{center}
\begin{tabular}{c|c||c|c}
\multicolumn{2}{c||}{Eulerian configuration} & \multicolumn{2}{c}{Lagrangian configuration}\\\hline Name &
Formula & Name & Formula \\\hline Hencky & $\ln{\mathbf V}$ & Hencky &$\ln{\mathbf U}$ \\ Swainger & ${\bf
1} -{\mathbf V}^{-1}$ & Biot &${\mathbf U}-{\mathbf 1}$ \\ Alamansi & $\frac{1}{2} ( {\mathbf 1} - {\mathbf V}^{-2})$ &
Green &$\frac{1}{2} ( {\mathbf U}^2 - {\mathbf 1})$
\end{tabular}
\caption{Eulerian and Lagrangian strains conjugate by the transformation tensor $\pmb{\mathcal A}$, {\it cf}.\,(\ref{e=A-Ta}b). \label{Tb}}
\end{center}
\end{table}

Assume further that a hyperelastic material satisfies the following strain energy function
\begin{equation}
\psi(\widehat{\boldsymbol\varepsilon})= \frac{1}{\widehat{\rho}} \bigg[ \frac{1}{2!} 
\widehat{c}_{i\mspace{-4.5mu}j\,k\mspace{-2.0mu}l}\,\widehat{\varepsilon}^{\,i\mspace{-4.5mu}j}\,\widehat{\varepsilon}^{\,k\mspace{-2.0mu}l}
+\frac{1}{3!}\widehat{C}_{i\mspace{-4.5mu}j\,k\mspace{-2.0mu}l\,m\mspace{-3.0mu}n}\,\widehat{\varepsilon}^{\,i\mspace{-4.5mu}j}\,\widehat{\varepsilon}^{\,k\mspace{-2.0mu}l}\,\widehat{\varepsilon}^{\,m\mspace{-3.0mu}n} 
\bigg], \label{psi2}
\end{equation}
where $\widehat{\mathbf c}$ and $\widehat{\mathbf C}$ are tensors of the second and third--order elastic coefficients
determined in relation to a given strain measure. Substitution of \eqref{psi2} into \eqref{ws} gives the following equation
for the conjugate stress
\begin{equation}
\widehat{\boldsymbol\sigma} = \widehat{\mathbf c}\!:\!\widehat{\boldsymbol\varepsilon}+
\frac12\,\widehat{\boldsymbol\varepsilon}\!:\!\widehat{\mathbf C}\!:\!\widehat{\boldsymbol\varepsilon} \,.
\end{equation}
Each of strain measures from the Seth-Hill family can be recalculated to another one from the same family by means of the
following transformation rule
\begin{equation}
\widehat{\boldsymbol\varepsilon}\,'(\widehat{\boldsymbol\varepsilon}) = \begin{cases}
\frac1{m'}\big[(m\,\widehat{\boldsymbol\varepsilon}+{\mathbf 1})^{\frac{m'}{m}} -{\mathbf 1} \big] & \text{for} \quad m\ne0 \land m'\ne 0\,,\\
\frac1{m'}\big[\exp{(m'\,\widehat{\boldsymbol\varepsilon})}-{\mathbf 1}\big] & \text{for} \quad m=0 \land m'\ne 0\,,\\
\frac1m\ln\big(m\,\widehat{\boldsymbol\varepsilon}+{\mathbf 1}) & \text{for} \quad m\ne0 \land m'=0 \,.
\end{cases} \label{ee'}
\end{equation}

It is easy to shown that SOECs are invariant with respect to the choice of strain measure. Contrary to that, TOECs depend
very strongly on the choice of strain measure. In other words, two experimenters observing the same run of instantaneous
stiffness under loading can determine two dramatically different sets of TOECs. Thus, in order to refer to the same initial
2nd order instantaneous stiffness by constitutive models based on different strain measures, TOECs must be recalculated from
the strain for which TOECs had been determined experimentally to the strain measure which is used in a given constitutive
model. The formulas for recalculation of TOECs have been given by \citet{Dlu00},
\begin{align}
\widehat{C}\,'_{i\mspace{-4.5mu}j\,k\mspace{-2.0mu}l\,m\mspace{-3.0mu}n}&= \widehat{C}_{i\mspace{-4.5mu}j\,k\mspace{-2.0mu}l\,m\mspace{-3.0mu}n} + (m-m') \big[
{\mathcal J}\!{_{i\mspace{-4.5mu}j\,k\mspace{-2.0mu}l}}^{\ae}\ \widehat{c}_{\ae\,m\mspace{-3.0mu}n} + 
{\mathcal J}\!{_{k\mspace{-2.0mu}l\,m\mspace{-3.0mu}n}}^{\ae}\ \widehat{c}_{\ae\,i\mspace{-4.5mu}j} +
{\mathcal J}\!{_{m\mspace{-3.0mu}n\,i\mspace{-4.5mu}j}}^{\ae}\ \widehat{c}_{\ae\,k\mspace{-2.0mu}l}\big] ,  \label{C'}
\end{align}  
where $\pmb{\mathcal{J}}$ is a sixth-order proper-symmetric unit tensor.

Apply \eqref{C'} to recalculation of TOECs. According to the method proposed by \citet{ThurstonS66}, the 2-nd order
instantaneous stiffness change of silicon crystal had been measured by \citet{JohalD06} and stored in the form of
TOECs related to Green strain measure, see Table \ref{3rdSi} . SOECs obtained by them were $\widehat{c}_{11}=166$\,GPa,
$\widehat{c}_{12}=64$\,GPa, $\widehat{c}_{44}= 80$\,GPa. In the case of cubic crystal, \eqref{C'} reads \cite{Dlu00}
\begin{align}
\widehat{C}_{111}' &= \widehat{C}_{111}+ 3           (m-m')\, \widehat{c}_{11} \label{C111} , &
\widehat{C}_{144}' &= \widehat{C}_{144}+ \frac{1}{2} (m-m')\, \widehat{c}_{12} ,\nonumber \\ 
\widehat{C}_{112}' &= \widehat{C}_{112}+             (m-m')\, \widehat{c}_{12}  , &
\widehat{C}_{155}' &= \widehat{C}_{155}+             (m-m')\Big[\widehat{c}_{44} +
\frac{1}{4}\widehat{c}_{12} +  \frac{1}{4} \widehat{c}_{11} \Big]   , \\
\widehat{C}_{123}' &= \widehat{C}_{123} , &
\widehat{C}_{456}' &= \widehat{C}_{456}+ \frac{3}{4} (m-m')\, \widehat{c}_{44} . \nonumber  
\end{align}
Thus, TOECs have been recalculated here onto equivalent TOECs referred to a few mutually different strain measures, see rows
3,4 and 5 in Table \ref{3rdSi}. Additionally, in the last row the strain measure has been chosen in such a way to get a
Hookean for which the second-order bulk modulus vanishes. Namely, the substitution of \ 
$\widehat{\mathbf c} = \widehat{\mathbf c}^{\:\circ} + \widehat{\mathbf C}\!:\!\widehat{\boldsymbol\varepsilon}$ \ into \ 
$\widehat{B} = \frac{\widehat{c}_{11} + 2 \widehat{c}_{12}}{3}$ gives the following dependence of bulk modulus on the volume
strain $\widehat{\varepsilon}$
\begin{align}
\widehat{B} (\,\widehat{\varepsilon}\,) = \frac{\widehat{c}^{\;\circ}_{11} +2\,\widehat{c}_{12}^{\;\circ}} 3 +
\frac{\widehat{C}_{111}+ 6\,\widehat{C}_{112} + 2\,\widehat{C}_{123}}3 \, \widehat{\varepsilon} \label{B2cubic}
\end{align}
where $\widehat{\varepsilon} =\widehat{\varepsilon}^{\,1} = \widehat{\varepsilon}^{\,2} = \widehat{\varepsilon}^{\,3}$.  One
can check that $\frac{d\,B} {d\,\widehat{\varepsilon}'}$ vanishes for $m'= m + \frac{\widehat{C}_{111}+6\,\widehat{C}_{112}+2\,\widehat{C}_{123}} {3\,\widehat{c}_{11}+6\,\widehat{c}_{12}}$. 
\begin{table}\begin{center}
\begin{tabular}{r|rrrrrrr|l}
$\widehat{\boldsymbol\varepsilon}=\frac1m({\mathbf U}^m-{\mathbf 1})$  & $\widehat{C}_{111}$ & $\widehat{C}_{112}$ & $\widehat{C}_{123}$ & $\widehat{C}_{144}$ & $\widehat{C}_{155}$ & $\widehat{C}_{456}$ & $\partial{\widehat{B}}/\partial{\widehat{\varepsilon}}$ & m  \\\hline 
Green  &  -815 & -450 & -75 &  16   & -307 & -82 &  -1124 &\ 2  \\  
Biot   &  -317 & -386 & -75 &  48   & -170 & -22 &   -928 &\ 1  \\  
Hencky &   181 & -322 & -75 &  80   &  -32 &  38 &   -544 &\ 0  \\
Almansi&  1177 & -194 & -75 & 144   &  243 & 158 &    -46 & -2   \\
       &  1254 & -184 & -75 & 149   &  266 & 167 &      0 & -2.155 
\end{tabular}
\caption{Silicon crystal third--order elastic coefficients [GPa] related to different strain measures. \label{3rdSi}} \end{center}
\end{table}

\subsection{Uniaxial stretch test} 
In this case the axial component of Hencky strain changes proportionally to the relative change of the
length of sample, according to the differential form
\begin{align}
d\,\widehat{\varepsilon} = d\,(\ln U ) = \frac{dl} l . \label{dll}
\end{align}
Bearing in mind \eqref{dll}, it is easy to note that
\begin{align}
\lim_{\Delta l \to 0}\frac{\Delta\sigma_{11}}{\frac{\Delta l}l}=\frac{d\,\sigma_{11}}{d(\ln F_{11})}&=\frac{d\,\sigma_{11}}{d\,F_{11}} \frac{d\,F_{11}}{d(\ln F_{11})}=\frac{d\,\sigma_{11}}{d\,F_{11}}F_{11}\ .
\end{align}
This means that the increase of Hencky strain is invariant with respect to the choice of the reference configuration. For
this reason, Hencky strain was initially regarded as a measure for which Hooke's law holds a constant instantaneous
stiffness. Nevertheless, the strain energy is referred to the mass density thus, also the instantaneous stiffness changes
during the uniaxial deformation process. The Hookean extended by TOECs term leads to the following relationship for the
Cauchy stress
\begin{equation}
\sigma = \begin{cases} 
c_\circ\,F^{m-1}\,\frac1 m (F^m-1)+ \frac12\, C_\circ F^{m-1} \frac1{m^2} (F^m -1)^2 & \text{for}\quad m\ne 0 ,\\
c_\circ\,F^{-1}\,\ln{F}+\frac12\, C_\circ\,F^{-1}\ln^2\!F & \text{for}\quad m = 0,
\end{cases}
\end{equation}
where $\sigma=\sigma_{11}$, $F=F_{11}$, $c_\circ=\widehat{c}_{1111}$ and $C_\circ=\widehat{C}_{111111}$. 
\begin{figure}
\begin{tikzpicture}[declare function={
f(\g,\m) = (((3*\m-1)*x^(3*\m)+(2-4*\m)*x^(2*\m)+(\m-1)*x^\m)*\g + (-9*\m^2+21*\m-6)*x^(3*\m) + (16*\m^2-32*\m+12)*x^(2*\m)
+ (-5*\m^2+11*\m-6)*x^\m)/(2*\m^2*x);}]
\begin{axis}[width=15.cm, height=5.5cm, ymin=0.2, 
minor tick num=1, ytick={0.1,0.2,0.4,0.6,0.8,1.0,1.2,1.4}, axis x line=middle, axis y line=left,
xlabel=$F_{11}$, ylabel = Instantaneous stiffness $/\,{\widehat{c}_{11}}$] 
\foreach \g in {-4.906} { 
\addplot[domain= 0.69:1.195,very thick,       loosely dotted, postaction={decorate}, decoration={raise=2pt, text along path, text={\ \ Green 2+3}}] {f(\g,2.)};
\addplot[domain=  0.7:1.075,  semithick,               dashed, postaction={decorate}, decoration={raise=2pt, text along path, text={Green 2\ }, text align={right}}] {f(0.,2.)};
\addplot[domain= 0.87:1.290,very thick,           dashdotted, postaction={decorate}, decoration={raise=2pt, text along path, text={Almansi 2+3},   text align={right}}] {f(\g,-2.)};
\addplot[domain=0.9565:1.25,  semithick,loosely dashdotdotted, postaction={decorate}, decoration={raise=2pt, text along path, text={{\,Almansi} 2},     text align={left}}]  {-(3*x^2-5)/(2*x^5)};
\addplot[domain= 0.69:1.45,  semithick,                       postaction={decorate}, decoration={raise=2pt, text along path, text={\ \ Hencky 2+3},text align={left}}]  {-((ln(x)^2-2*ln(x))*\g+6*ln(x)^2-10*ln(x)-2)/(2*x)};
\addplot[domain= 0.85:1.45,      thick,               dashed, postaction={decorate}, decoration={raise=2pt, text along path, text={Hencky 2\ \ \ },text align={right}}] {-(ln(x)-1)/x};
};
\end{axis}
\end{tikzpicture}
\begin{tikzpicture}[declare function={
f(\g,\m) = ((x^(3*\m)-2*x^(2*\m)+x^\m)*\g+(6-3*\m)*x^(3*\m)+(8*\m-12)*x^(2*\m)+(6-5*\m)*x^\m)/(2*\m^2*x);}]
\begin{axis}[width=15.cm, height=5.5cm, minor tick num=1, ymax=0.32, axis y line=left, axis x line=middle,
xlabel=$F_{11}$, ylabel = Cauchy stress $/\,\widehat{c}_{11}$] 
\foreach \g in {-4.906} { 
\addplot[domain=0.69:1.24 ,very thick, loosely dotted, postaction={decorate}, decoration={raise=2pt, text along path, text={\ \ Green 2+3}}] {f(\g,2.)};
\addplot[domain=0.7:1.20,   semithick, dashed, postaction={decorate}, decoration={raise=2pt, text along path, text={Green 2}, text align={right}}] {f(0.,2.)};
\addplot[domain=0.84:1.45, very thick,dashdotted, postaction={decorate}, decoration={raise=1pt, text along path, text={Almansi 2+3}, text align={right}}] {f(\g,-2.)};
\addplot[domain=0.83:1.375, semithick,loosely dashdotdotted, postaction={decorate}, decoration={raise=1pt, text along path, text={Almansi 2}, text align={right}}] {(x^2-1)/(2*x^5)};
\addplot[domain=0.69:1.325, semithick, postaction={decorate}, decoration={raise=2pt, text along path, text={Hencky 2+3}, text align={right}}] {(ln(x)^2*\g+6*ln(x)^2+2*ln(x))/(2*x)};
\addplot[domain=0.74:1.45, very thick, dashed, postaction={decorate}, decoration={raise=2pt, text along path, text={Hencky 2}, text align={right}}] {ln(x)/x};
};
\end{axis}
\end{tikzpicture}
\caption{Instantaneous stiffness and Cauchy stress curves for constitutive models obtained by adding third--order elastic
  terms to Hookean models in uniaxial stretch test, {\it cf}.\,Table\,\ref{3rdSi}.\label{sethfig}}
\end{figure}
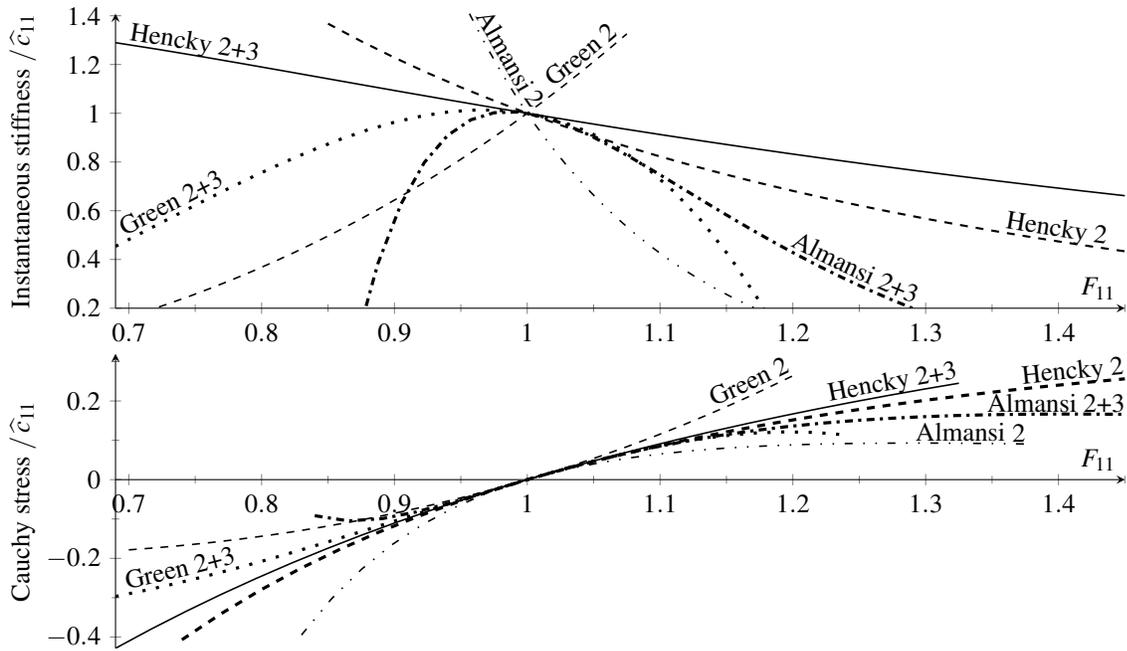
In order to hold the same 2nd order instantaneous stiffness in the relaxed configuration by different constitutive models,
the TOECs must be recalculated first to the values corresponding to strain measure used in the given constitutive
model. From a mathematical point of view, the instantaneous stiffness tensor is identified with the derivative of Cauchy
stress over the deformation gradient in the current configuration assumed to be the reference one, see \cite{Ogden84}.
As an example, consider the behaviour of the 2nd and 3rd order hyperelastic models of silicon crystal. In contrast to SOECs,
the TOECs have been recalculated according to the state-of-the-art discussed above, see \eqref{C'} and Table
1. Thanks to such a chosen transformation rule, all of the third--order elastic models hold the same, experimentally
observed change of {\em instantaneous} stiffness in the vicinity of the relaxed configuration. In other words, despite of
the different TOECs used, the Cauchy stress and instantaneous stiffness diagrams shown in Fig.\,\ref{sethfig} for
Green\,2+3, Hencky\,2+3, Almansi\,2+3, have the same initial slope corresponding to the experimentally observed
instantaneous stiffness evolution, {\it cf}.\,$\frac{d^2\sigma}{d\,F_{11}^2}\big|_{F_{11}=1}$.  All runs shown have been
normalized by dividing both the axial Cauchy stress and instantaneous stiffness by $\widehat{c}_{11}$.

It is worth emphasizing that a one-dimensional Hookean model of atomic bound based on the use of Hencky strain demonstrates
the constant stiffness of atomic bond. As is shown in solid physics the stiffness of atomic bonds decrease together
with stretching the bonds. Obviously, these two effects, the volume change vs the change of single atomic bonds stiffness,
having two mutually different origin, contribute together in the overall response of elastic material as a whole. It was
shown in many papers that Hookean based on Hencky strain pretty well approximates the atomistic simulations for many
monocrystals, see e.g.\ \cite{DluT03,Mazdziarz15}.

\section{TOECs in finite element modelling}
In many iterative schemes applied in solving nonlinear FE problems the important role takes the correct determination of the
tangent stiffness matrix. In order to limit considerations to matrix components resulting from constitutive equations
discussed above, we consider the classical quasi-static problem. In such a case, the total potential energy $\Pi$ is defined
as
\begin{align}
\Pi &= \int_V \widehat{\rho}\, \psi \det{\mathbf F}^{-1} dv - \Pi_{\rm ext} , \label{Pi}
\end{align}
where $\Pi_{\rm ext}=\int_{\partial V} {\mathbf u}\, \boldsymbol\sigma d{\mathbf s} $ denotes the potential of external
traction forces acting on the surface $\partial V$ bounding the volume region V.  In the case of the isoparametric finite
elements, the virtual displacement field is determined by the following relation:
$\delta u^i = \sum_{\sf m} N_{\sf mi}\,\delta {a_{\sf m}}^i$, where $N_{\sf mi}$ and $\delta {a_{\sf m}}^i$ denote the shape
function and a variation of $i$-th displacement component at node $\sf m$, respectively. For isoparametric elements, the
same shape function is assumed for all spatial position vectors and displacement vectors, i.e.
$N_{\sf m1}({\mathbf a}_1,\ldots,{\mathbf a}_{\sf n})= N_{\sf m2}({\mathbf a}_1,\ldots,{\mathbf a}_{\sf n}) = N_{\sf
  m3}({\mathbf a}_1,\ldots,{\mathbf a}_{\sf n})= N_{\sf m}({\mathbf a}_1,\ldots,{\mathbf a}_{\sf n})$.  Nevertheless, to
distinguish which shape function and gradients relate to which nodal displacement component, we save in our notation the
displacement component index and write it in serif letters contrary to italics denoting the co-- and contravariant
components of vectors and tensors, respectively. The variation of total energy \eqref{Pi} can be rewritten as
\begin{align}
\delta \Pi & = \delta {a_{\sf m}}^i\int_V\widehat{\rho}\,\fracp{\psi}{\widehat{\varepsilon}_{IJ}} \fracp{\widehat{\varepsilon}_{IJ}} {{a_{\sf m}}^i} \det{\mathbf F}^{-1} {\rm d}v - 
\delta {a_{\sf m}}^i \int_{\partial V} N_{\sf mi}\, \sigma_{ij} n^j{\rm d}s .\label{varP}
\end{align}
where
\begin{align}
\fracp{\widehat{\varepsilon}_{KL}} {{a_{\sf m}}^i} 
& = \frac12\,{{\mathcal A}_{KL}}^{kl}\,\big(g_{ik}\, N_{{\sf mi},l} + g_{il}\, N_{{\sf mi},k}\big)\,.  \label{deda1}
\end{align}
Thus, the elimination of virtual displacements from \eqref{varP} leads to the following nonlinear equation
\begin{align}
P_{{\sf m}\,i} &\stackrel{df}= \frac 1 2 \int_V \big(g_{ik}\, N_{{\sf mi},l} + g_{il}\, N_{{\sf mi},k}\big)
{\sigma}^{kl} {\rm d}v -\int_{\partial V} N_{\sf mi}\, \sigma_{ij}\, n^j\,{\rm d}s = 0 . \label{Pmi}
\end{align}
where $P_{{\sf m}\,i}$ is the $i$-th component of the residual vector in the $\sf m$-th node. As shown in Appendix, such
obtained tangent stiffness matrix takes form
\begin{gather}
K{_{\sf m}}^i{_{\sf n}}^j 
= \int_{V}\!\!  N_{{\sf mi},k} \Big[\Big({\mathcal A}{_{PR}}^{ik}
\fracp{\widehat{\sigma}^{PR}}{\widehat{\varepsilon}_{ST}} {{\mathcal A}_{ST}}^{jl} +  
{\mathcal B}{_{PR}}^{ikjl}\, \widehat{\sigma}^{PR}\Big) \det{\mathbf F}^{-1}  +   g^{ij} \sigma^{kl}\Big] N_{{\sf nj},l}\, dv . \label{Kij}
\end{gather}

\subsection{Numerical example}
\begin{figure}
\begin{center}
\includegraphics[height=7.cm]{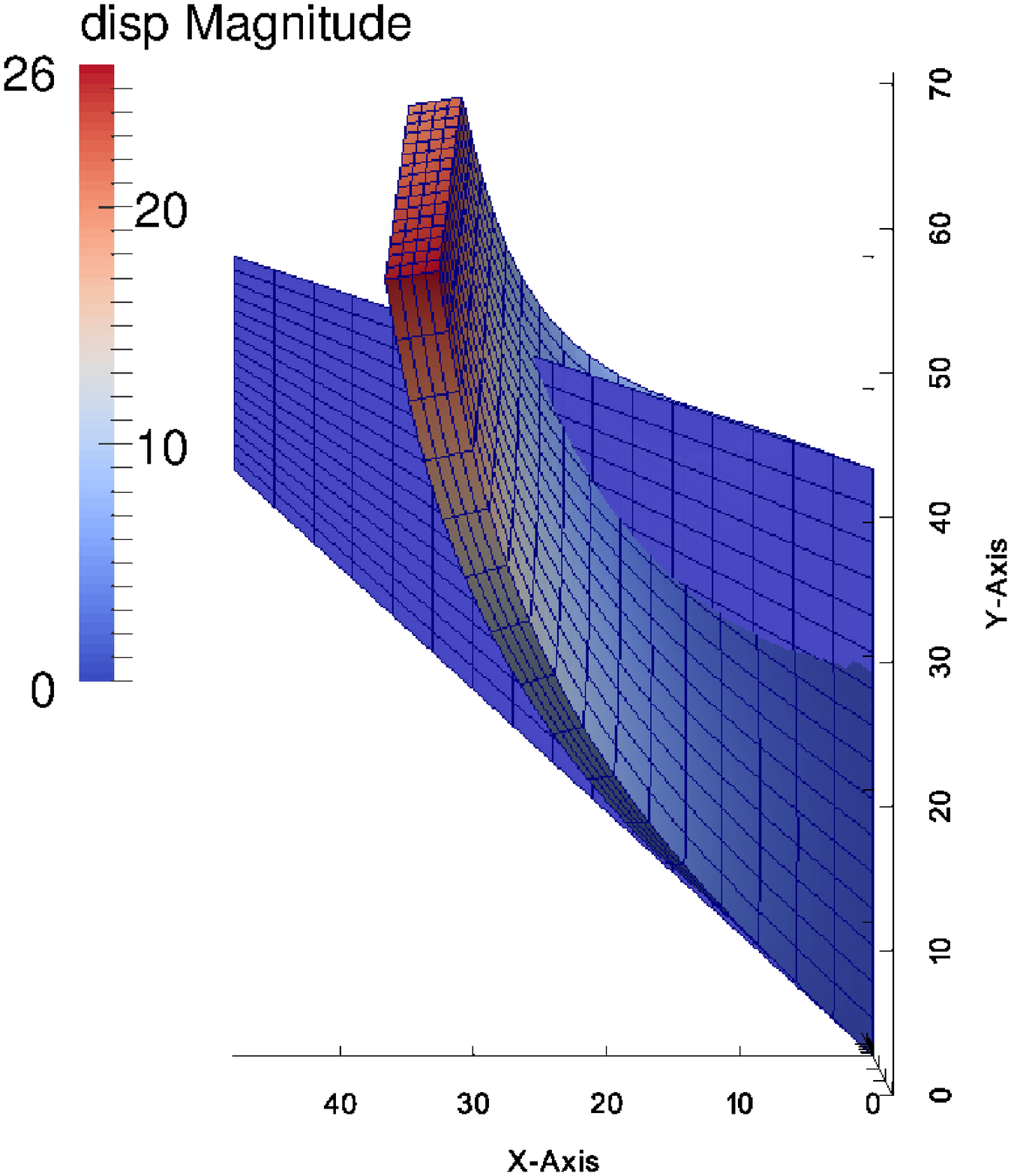} 
\includegraphics[height=7.cm]{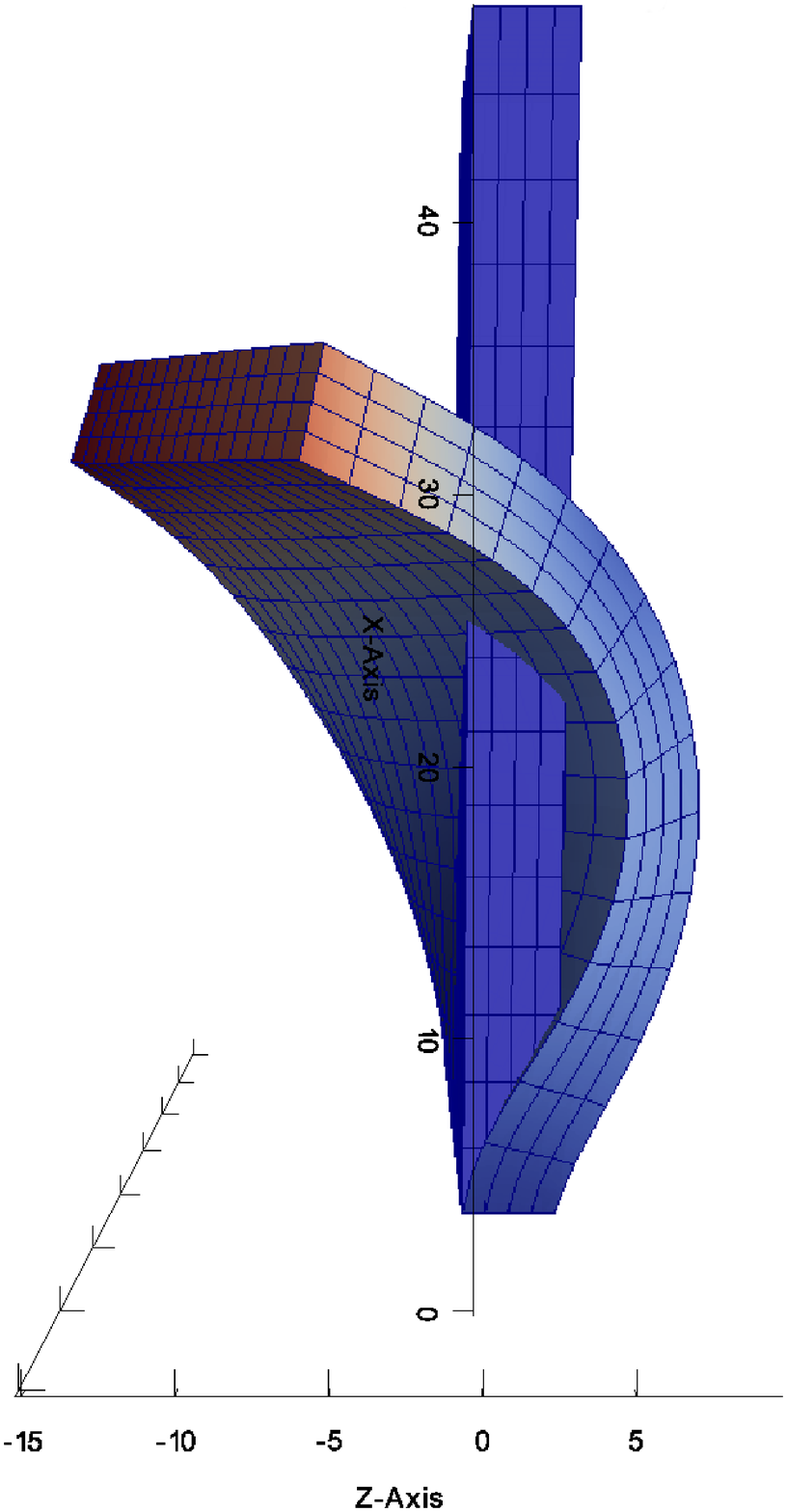} 
\end{center}
\caption{Cook{'}s membrane problem. The initial and final configurations for the Hencky\,2+3 model and
   Si nanomembrane orientation $[075](3\bar57)$.} \label{fig:parview}
\end{figure}

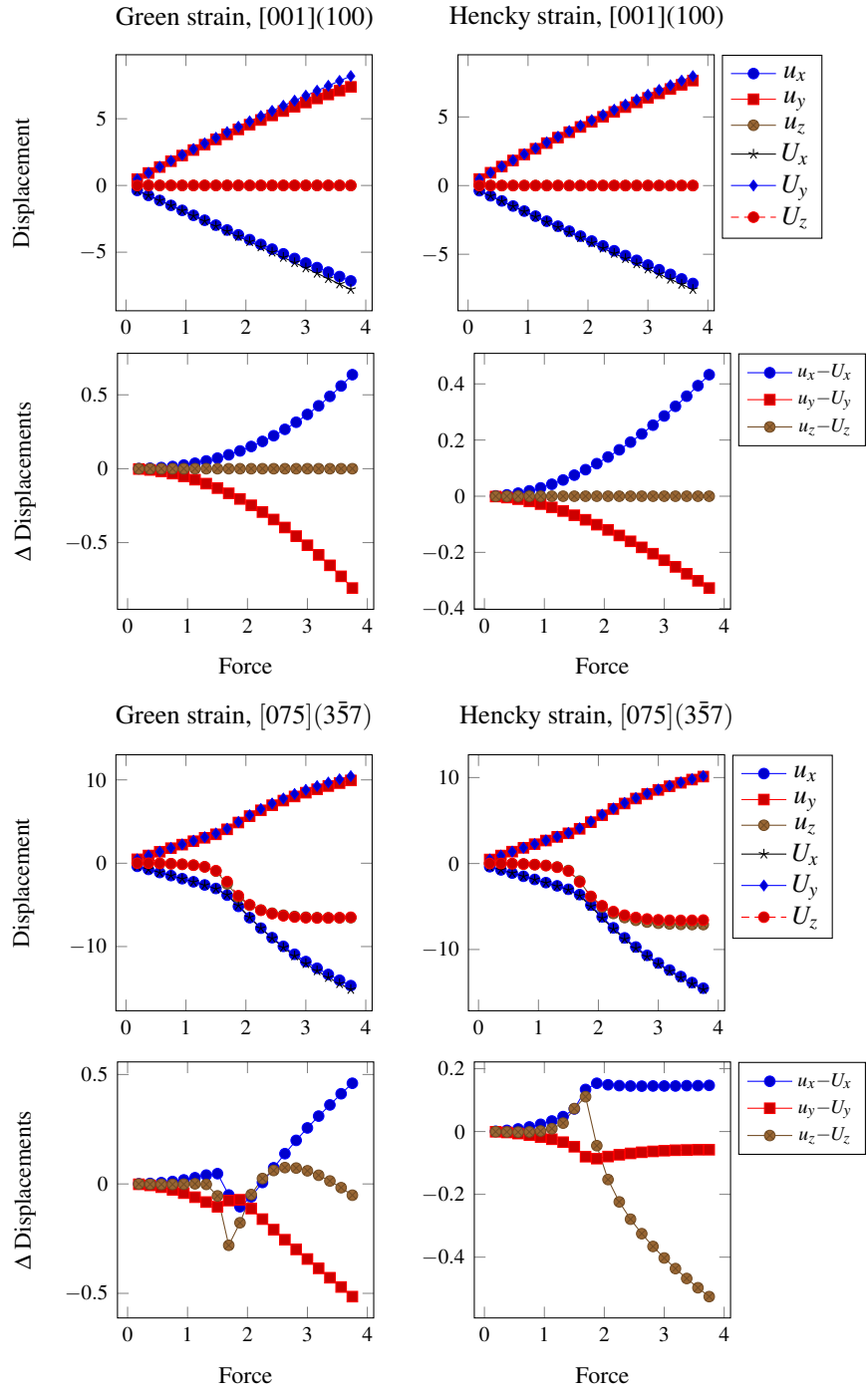
\begin{figure}
\begin{center}
\pgfplotsset{small}
\begin{tabular}{rl}
\begin{tikzpicture}
\begin{axis}[baseline, trim axis left,title={Green strain, [001](100)},width=5.cm, height=5.cm, ylabel={Displacement}]
\addplot coordinates{
(0.1875, -3.7317E-01)
(0.3750, -7.4671E-01)
(0.5625, -1.1202E+00)
(0.7500, -1.4931E+00)
(0.9375, -1.8652E+00)
(1.1250, -2.2360E+00)
(1.3125, -2.6053E+00)
(1.5000, -2.9726E+00)
(1.6875, -3.3377E+00)
(1.8750, -3.7004E+00)
(2.0625, -4.0604E+00)
(2.2500, -4.4175E+00)
(2.4375, -4.7716E+00)
(2.6250, -5.1224E+00)
(2.8125, -5.4698E+00)
(3.0000, -5.8138E+00)
(3.1875, -6.1541E+00)
(3.3750, -6.4908E+00)
(3.5625, -6.8236E+00)
(3.7500, -7.1527E+00)};

\addplot coordinates{
(0.1875, 4.7213E-01)
(0.3750, 9.3202E-01)
(0.5625, 1.3799E+00)
(0.7500, 1.8158E+00)
(0.9375, 2.2402E+00)
(1.1250, 2.6531E+00)
(1.3125, 3.0548E+00)
(1.5000, 3.4457E+00)
(1.6875, 3.8259E+00)
(1.8750, 4.1958E+00)
(2.0625, 4.5556E+00)
(2.2500, 4.9056E+00)
(2.4375, 5.2462E+00)
(2.6250, 5.5776E+00)
(2.8125, 5.9001E+00)
(3.0000, 6.2140E+00)
(3.1875, 6.5196E+00)
(3.3750, 6.8172E+00)
(3.5625, 7.1071E+00)
(3.7500, 7.3894E+00)};

\addplot coordinates{
(0.1875,  3.4969E-15)
(0.3750,  1.0095E-16)
(0.5625, -7.6802E-15)
(0.7500, -7.0016E-16)
(0.9375, -1.6714E-15)
(1.1250,  2.7038E-16)
(1.3125, -5.9842E-15)
(1.5000, -2.9359E-15)
(1.6875, -6.9248E-15)
(1.8750,  4.9854E-16)
(2.0625,  2.5290E-15)
(2.2500,  8.9660E-15)
(2.4375,  2.5993E-15)
(2.6250, -2.0924E-16)
(2.8125, -3.6092E-15)
(3.0000, -6.9976E-15)
(3.1875, -1.0565E-14)
(3.3750, -1.3210E-16)
(3.5625,  1.2555E-14)
(3.7500,  3.3701E-15)};

\addplot coordinates{
(0.1875,-3.7396E-01)
(0.3750,-7.5007E-01)
(0.5625,-1.1282E+00)
(0.7500,-1.5082E+00)
(0.9375,-1.8899E+00)
(1.1250,-2.2732E+00)
(1.3125,-2.6580E+00)
(1.5000,-3.0442E+00)
(1.6875,-3.4318E+00)
(1.8750,-3.8207E+00)
(2.0625,-4.2108E+00)
(2.2500,-4.6022E+00)
(2.4375,-4.9948E+00)
(2.6250,-5.3888E+00)
(2.8125,-5.7842E+00)
(3.0000,-6.1812E+00)
(3.1875,-6.5798E+00)
(3.3750,-6.9805E+00)
(3.5625,-7.3835E+00)
(3.7500,-7.7891E+00)};      

\addplot coordinates{
(0.1875, 4.7418E-01)
(0.3750, 9.4025E-01)
(0.5625, 1.3984E+00)
(0.7500, 1.8488E+00)
(0.9375, 2.2917E+00)
(1.1250, 2.7272E+00)
(1.3125, 3.1557E+00)
(1.5000, 3.5772E+00)
(1.6875, 3.9921E+00)
(1.8750, 4.4005E+00)
(2.0625, 4.8028E+00)
(2.2500, 5.1992E+00)
(2.4375, 5.5899E+00)
(2.6250, 5.9754E+00)
(2.8125, 6.3558E+00)
(3.0000, 6.7316E+00)
(3.1875, 7.1031E+00)
(3.3750, 7.4708E+00)
(3.5625, 7.8351E+00)
(3.7500, 8.1966E+00)};

\addplot coordinates{
(0.1875,-3.6787E-15)
(0.3750,-7.5723E-16)
(0.5625, 1.6598E-16)
(0.7500,-7.1622E-16)
(0.9375,-6.5382E-16)
(1.1250,-2.8024E-15)
(1.3125, 9.0733E-16)
(1.5000, 1.7743E-14)
(1.6875,-1.4449E-14)
(1.8750, 6.7692E-15)
(2.0625, 3.4337E-15)
(2.2500,-4.0684E-15)
(2.4375,-2.4224E-15)
(2.6250,-7.1729E-15)
(2.8125, 2.3046E-15)
(3.0000,-3.7421E-15)
(3.1875, 9.6198E-16)
(3.3750,-8.7069E-15)
(3.5625, 4.5950E-15)
(3.7500, 2.2940E-15)};

\end{axis}
\end{tikzpicture}&
\begin{tikzpicture}
\begin{axis}[baseline, trim axis left, title={Hencky strain, [001](100)}, legend pos = outer north east, width=5.cm, height=5.cm]
\addplot coordinates{
(0.1875,-3.7274E-01)
(0.3750,-7.4511E-01)
(0.5625,-1.1168E+00)
(0.7500,-1.4876E+00)
(0.9375,-1.8572E+00)
(1.1250,-2.2254E+00)
(1.3125,-2.5920E+00)
(1.5000,-2.9567E+00)
(1.6875,-3.3193E+00)
(1.8750,-3.6797E+00)
(2.0625,-4.0376E+00)
(2.2500,-4.3930E+00)
(2.4375,-4.7457E+00)
(2.6250,-5.0955E+00)
(2.8125,-5.4423E+00)
(3.0000,-5.7861E+00)
(3.1875,-6.1267E+00)
(3.3750,-6.4640E+00)
(3.5625,-6.7980E+00)
(3.7500,-7.1285E+00)};
       
\addplot coordinates{
(0.1875, 4.7292E-01)
(0.3750, 9.3520E-01)
(0.5625, 1.3870E+00)
(0.7500, 1.8284E+00)
(0.9375, 2.2597E+00)
(1.1250, 2.6810E+00)
(1.3125, 3.0926E+00)
(1.5000, 3.4945E+00)
(1.6875, 3.8870E+00)
(1.8750, 4.2704E+00)
(2.0625, 4.6448E+00)
(2.2500, 5.0104E+00)
(2.4375, 5.3675E+00)
(2.6250, 5.7163E+00)
(2.8125, 6.0569E+00)
(3.0000, 6.3897E+00)
(3.1875, 6.7148E+00)
(3.3750, 7.0325E+00)
(3.5625, 7.3429E+00)
(3.7500, 7.6462E+00)};
       
\addplot coordinates{
(0.1875, 1.4067E-15)
(0.3750,-1.6357E-15)
(0.5625, 1.1189E-15)
(0.7500, 6.6382E-15)
(0.9375,-6.2334E-15)
(1.1250, 1.1167E-15)
(1.3125, 3.1316E-15)
(1.5000, 6.2983E-15)
(1.6875, 5.4382E-14)
(1.8750,-4.3548E-15)
(2.0625,-1.8754E-15)
(2.2500,-1.7527E-15)
(2.4375,-6.5333E-15)
(2.6250, 3.4520E-15)
(2.8125,-5.2197E-15)
(3.0000,-9.8233E-15)
(3.1875, 3.2152E-15)
(3.3750, 2.8207E-15)
(3.5625, 2.0604E-15)
(3.7500, 9.4557E-16)};
       
\addplot coordinates{
(0.1875,-3.7393E-01)
(0.3750,-7.4989E-01)
(0.5625,-1.1276E+00)
(0.7500,-1.5067E+00)
(0.9375,-1.8869E+00)
(1.1250,-2.2680E+00)
(1.3125,-2.6497E+00)
(1.5000,-3.0317E+00)
(1.6875,-3.4138E+00)
(1.8750,-3.7957E+00)
(2.0625,-4.1772E+00)
(2.2500,-4.5582E+00)
(2.4375,-4.9383E+00)
(2.6250,-5.3174E+00)
(2.8125,-5.6953E+00)
(3.0000,-6.0718E+00)
(3.1875,-6.4468E+00)
(3.3750,-6.8202E+00)
(3.5625,-7.1917E+00)
(3.7500,-7.5613E+00)};
       
\addplot coordinates{
(0.1875, 4.7415E-01)
(0.3750, 9.4000E-01)
(0.5625, 1.3976E+00)
(0.7500, 1.8469E+00)
(0.9375, 2.2879E+00)
(1.1250, 2.7208E+00)
(1.3125, 3.1455E+00)
(1.5000, 3.5622E+00)
(1.6875, 3.9710E+00)
(1.8750, 4.3718E+00)
(2.0625, 4.7649E+00)
(2.2500, 5.1502E+00)
(2.4375, 5.5281E+00)
(2.6250, 5.8985E+00)
(2.8125, 6.2617E+00)
(3.0000, 6.6177E+00)
(3.1875, 6.9667E+00)
(3.3750, 7.3088E+00)
(3.5625, 7.6443E+00)
(3.7500, 7.9732E+00)};
       
\addplot coordinates{
(0.1875,-3.0496E-15)
(0.3750,-1.6256E-15)
(0.5625,-4.1373E-16)
(0.7500, 3.4123E-15)
(0.9375,-2.2730E-15)
(1.1250, 2.1213E-15)
(1.3125,-7.3454E-15)
(1.5000,-4.3989E-15)
(1.6875,-1.4005E-14)
(1.8750,-1.3985E-15)
(2.0625,-1.5202E-15)
(2.2500, 7.3934E-16)
(2.4375,-6.9635E-16)
(2.6250,-6.0217E-15)
(2.8125, 4.8080E-15)
(3.0000,-7.6005E-15)
(3.1875,-3.1902E-15)
(3.3750, 6.8202E-16)
(3.5625, 2.7946E-16)
(3.7500,-1.7066E-15)};

\legend{$u_x$,$u_y$,$u_z$,$U_x$,$U_y$,$U_z$} 
\end{axis}
\end{tikzpicture}\\
\begin{tikzpicture}
\begin{axis}[baseline, trim axis left, ylabel= $\Delta$ Displacements, width=5.cm, height=5.cm, xlabel= Force]
\addplot coordinates{
(0.1875, -3.7317E-01+3.7396E-01)
(0.3750, -7.4671E-01+7.5007E-01)
(0.5625, -1.1202E+00+1.1282E+00)
(0.7500, -1.4931E+00+1.5082E+00)
(0.9375, -1.8652E+00+1.8899E+00)
(1.1250, -2.2360E+00+2.2732E+00)
(1.3125, -2.6053E+00+2.6580E+00)
(1.5000, -2.9726E+00+3.0442E+00)
(1.6875, -3.3377E+00+3.4318E+00)
(1.8750, -3.7004E+00+3.8207E+00)
(2.0625, -4.0604E+00+4.2108E+00)
(2.2500, -4.4175E+00+4.6022E+00)
(2.4375, -4.7716E+00+4.9948E+00)
(2.6250, -5.1224E+00+5.3888E+00)
(2.8125, -5.4698E+00+5.7842E+00)
(3.0000, -5.8138E+00+6.1812E+00)
(3.1875, -6.1541E+00+6.5798E+00)
(3.3750, -6.4908E+00+6.9805E+00)
(3.5625, -6.8236E+00+7.3835E+00)
(3.7500, -7.1527E+00+7.7891E+00)};
 
\addplot coordinates{
(0.1875, 4.7213E-01-4.7418E-01)
(0.3750, 9.3202E-01-9.4025E-01)
(0.5625, 1.3799E+00-1.3984E+00)
(0.7500, 1.8158E+00-1.8488E+00)
(0.9375, 2.2402E+00-2.2917E+00)
(1.1250, 2.6531E+00-2.7272E+00)
(1.3125, 3.0548E+00-3.1557E+00)
(1.5000, 3.4457E+00-3.5772E+00)
(1.6875, 3.8259E+00-3.9921E+00)
(1.8750, 4.1958E+00-4.4005E+00)
(2.0625, 4.5556E+00-4.8028E+00)
(2.2500, 4.9056E+00-5.1992E+00)
(2.4375, 5.2462E+00-5.5899E+00)
(2.6250, 5.5776E+00-5.9754E+00)
(2.8125, 5.9001E+00-6.3558E+00)
(3.0000, 6.2140E+00-6.7316E+00)
(3.1875, 6.5196E+00-7.1031E+00)
(3.3750, 6.8172E+00-7.4708E+00)
(3.5625, 7.1071E+00-7.8351E+00)
(3.7500, 7.3894E+00-8.1966E+00)};
 
\addplot coordinates{
(0.1875,  3.4969E-15+3.6787E-15)
(0.3750,  1.0095E-16+7.5723E-16)
(0.5625, -7.6802E-15-1.6598E-16)
(0.7500, -7.0016E-16+7.1622E-16)
(0.9375, -1.6714E-15+6.5382E-16)
(1.1250,  2.7038E-16+2.8024E-15)
(1.3125, -5.9842E-15-9.0733E-16)
(1.5000, -2.9359E-15-1.7743E-14)
(1.6875, -6.9248E-15+1.4449E-14)
(1.8750,  4.9854E-16-6.7692E-15)
(2.0625,  2.5290E-15-3.4337E-15)
(2.2500,  8.9660E-15+4.0684E-15)
(2.4375,  2.5993E-15+2.4224E-15)
(2.6250, -2.0924E-16+7.1729E-15)
(2.8125, -3.6092E-15-2.3046E-15)
(3.0000, -6.9976E-15+3.7421E-15)
(3.1875, -1.0565E-14-9.6198E-16)
(3.3750, -1.3210E-16+8.7069E-15)
(3.5625,  1.2555E-14-4.5950E-15)
(3.7500,  3.3701E-15-2.2940E-15)};
 
\end{axis}
\end{tikzpicture}&
\begin{tikzpicture}
         \begin{axis}[baseline, trim axis left, width=5.cm, height=5.cm, legend pos = outer north east, xlabel= Force]
\addplot coordinates{
(0.1875, -3.7274E-01+3.7393E-01)
(0.3750, -7.4511E-01+7.4989E-01)
(0.5625, -1.1168E+00+1.1276E+00)
(0.7500, -1.4876E+00+1.5067E+00)
(0.9375, -1.8572E+00+1.8869E+00)
(1.1250, -2.2254E+00+2.2680E+00)
(1.3125, -2.5920E+00+2.6497E+00)
(1.5000, -2.9567E+00+3.0317E+00)
(1.6875, -3.3193E+00+3.4138E+00)
(1.8750, -3.6797E+00+3.7957E+00)
(2.0625, -4.0376E+00+4.1772E+00)
(2.2500, -4.3930E+00+4.5582E+00)
(2.4375, -4.7457E+00+4.9383E+00)
(2.6250, -5.0955E+00+5.3174E+00)
(2.8125, -5.4423E+00+5.6953E+00)
(3.0000, -5.7861E+00+6.0718E+00)
(3.1875, -6.1267E+00+6.4468E+00)
(3.3750, -6.4640E+00+6.8202E+00)
(3.5625, -6.7980E+00+7.1917E+00)
(3.7500, -7.1285E+00+7.5613E+00)};
 
\addplot coordinates{
(0.1875, 4.7292E-01-4.7415E-01)
(0.3750, 9.3520E-01-9.4000E-01)
(0.5625, 1.3870E+00-1.3976E+00)
(0.7500, 1.8284E+00-1.8469E+00)
(0.9375, 2.2597E+00-2.2879E+00)
(1.1250, 2.6810E+00-2.7208E+00)
(1.3125, 3.0926E+00-3.1455E+00)
(1.5000, 3.4945E+00-3.5622E+00)
(1.6875, 3.8870E+00-3.9710E+00)
(1.8750, 4.2704E+00-4.3718E+00)
(2.0625, 4.6448E+00-4.7649E+00)
(2.2500, 5.0104E+00-5.1502E+00)
(2.4375, 5.3675E+00-5.5281E+00)
(2.6250, 5.7163E+00-5.8985E+00)
(2.8125, 6.0569E+00-6.2617E+00)
(3.0000, 6.3897E+00-6.6177E+00)
(3.1875, 6.7148E+00-6.9667E+00)
(3.3750, 7.0325E+00-7.3088E+00)
(3.5625, 7.3429E+00-7.6443E+00)
(3.7500, 7.6462E+00-7.9732E+00)};
 
\addplot coordinates{
(0.1875,  1.4067E-15+3.0496E-15)
(0.3750, -1.6357E-15+1.6256E-15)
(0.5625,  1.1189E-15+4.1373E-16)
(0.7500,  6.6382E-15-3.4123E-15)
(0.9375, -6.2334E-15+2.2730E-15)
(1.1250,  1.1167E-15-2.1213E-15)
(1.3125,  3.1316E-15+7.3454E-15)
(1.5000,  6.2983E-15+4.3989E-15)
(1.6875,  5.4382E-14+1.4005E-14)
(1.8750, -4.3548E-15+1.3985E-15)
(2.0625, -1.8754E-15+1.5202E-15)
(2.2500, -1.7527E-15-7.3934E-16)
(2.4375, -6.5333E-15+6.9635E-16)
(2.6250,  3.4520E-15+6.0217E-15)
(2.8125, -5.2197E-15-4.8080E-15)
(3.0000, -9.8233E-15+7.6005E-15)
(3.1875,  3.2152E-15+3.1902E-15)
(3.3750,  2.8207E-15-6.8202E-16)
(3.5625,  2.0604E-15-2.7946E-16)
(3.7500,  9.4557E-16+1.7066E-15)};

\legend{$\scriptstyle u_x-U_x$,$\scriptstyle u_y-U_y$,$\scriptstyle u_z - U_z$}
\end{axis}
\end{tikzpicture}\\
\begin{tikzpicture}
\begin{axis}[baseline, trim axis left, title={Green strain, $[075](3\bar 57)$}, width=5.cm, height=5.cm, ylabel={Displacement}]
\addplot coordinates{
(0.1875,-3.7104E-01)
(0.3750,-7.4247E-01)
(0.5625,-1.1139E+00)
(0.7500,-1.4852E+00)
(0.9375,-1.8565E+00)
(1.1250,-2.2286E+00)
(1.3125,-2.6071E+00)
(1.5000,-3.0302E+00)
(1.6875,-3.8338E+00)
(1.8750,-5.1801E+00)
(2.0625,-6.5388E+00)
(2.2500,-7.7926E+00)
(2.4375,-8.9349E+00)
(2.6250,-9.9757E+00)
(2.8125,-1.0927E+01)
(3.0000,-1.1800E+01)
(3.1875,-1.2605E+01)
(3.3750,-1.3351E+01)
(3.5625,-1.4043E+01)
(3.7500,-1.4690E+01)};
       
\addplot coordinates{
(0.1875, 4.6919E-01)
(0.3750, 9.2646E-01)
(0.5625, 1.3721E+00)
(0.7500, 1.8063E+00)
(0.9375, 2.2296E+00)
(1.1250, 2.6432E+00)
(1.3125, 3.0502E+00)
(1.5000, 3.4711E+00)
(1.6875, 4.0668E+00)
(1.8750, 4.8832E+00)
(2.0625, 5.6589E+00)
(2.2500, 6.3507E+00)
(2.4375, 6.9657E+00)
(2.6250, 7.5159E+00)
(2.8125, 8.0119E+00)
(3.0000, 8.4627E+00)
(3.1875, 8.8756E+00)
(3.3750, 9.2565E+00)
(3.5625, 9.6100E+00)
(3.7500, 9.9401E+00)};
       
\addplot coordinates{
(0.1875, 1.1556E-02)
(0.3750, 3.0172E-03)
(0.5625,-2.1228E-02)
(0.7500,-6.2343E-02)
(0.9375,-1.2705E-01)
(1.1250,-2.3339E-01)
(1.3125,-4.3481E-01)
(1.5000,-9.4838E-01)
(1.6875,-2.4837E+00)
(1.8750,-4.0756E+00)
(2.0625,-5.0323E+00)
(2.2500,-5.6214E+00)
(2.4375,-5.9964E+00)
(2.6250,-6.2372E+00)
(2.8125,-6.3898E+00)
(3.0000,-6.4821E+00)
(3.1875,-6.5325E+00)
(3.3750,-6.5530E+00)
(3.5625,-6.5519E+00)
(3.7500,-6.5352E+00)};
       
\addplot coordinates{
(0.1875,-3.7159E-01)
(0.3750,-7.4485E-01)
(0.5625,-1.1197E+00)
(0.7500,-1.4963E+00)
(0.9375,-1.8749E+00)
(1.1250,-2.2569E+00)
(1.3125,-2.6475E+00)
(1.5000,-3.0771E+00)
(1.6875,-3.7835E+00)
(1.8750,-5.0759E+00)
(2.0625,-6.4801E+00)
(2.2500,-7.7992E+00)
(2.4375,-9.0087E+00)
(2.6250,-1.0114E+01)
(2.8125,-1.1126E+01)
(3.0000,-1.2056E+01)
(3.1875,-1.2915E+01)
(3.3750,-1.3712E+01)
(3.5625,-1.4455E+01)
(3.7500,-1.5150E+01)};
       
\addplot coordinates{
(0.1875, 4.7083E-01)
(0.3750, 9.3307E-01)
(0.5625, 1.3870E+00)
(0.7500, 1.8330E+00)
(0.9375, 2.2716E+00)
(1.1250, 2.7038E+00)
(1.3125, 3.1330E+00)
(1.5000, 3.5750E+00)
(1.6875, 4.1430E+00)
(1.8750, 4.9554E+00)
(2.0625, 5.7717E+00)
(2.2500, 6.5115E+00)
(2.4375, 7.1744E+00)
(2.6250, 7.7707E+00)
(2.8125, 8.3114E+00)
(3.0000, 8.8058E+00)
(3.1875, 9.2616E+00)
(3.3750, 9.6851E+00)
(3.5625, 1.0081E+01)
(3.7500, 1.0455E+01)};
       
\addplot coordinates{
(0.1875, 1.2048E-02)
(0.3750, 4.3945E-03)
(0.5625,-1.9161E-02)
(0.7500,-6.0230E-02)
(0.9375,-1.2587E-01)
(1.1250,-2.3384E-01)
(1.3125,-4.3263E-01)
(1.5000,-8.9371E-01)
(1.6875,-2.2033E+00)
(1.8750,-3.8982E+00)
(2.0625,-4.9838E+00)
(2.2500,-5.6461E+00)
(2.4375,-6.0577E+00)
(2.6250,-6.3118E+00)
(2.8125,-6.4623E+00)
(3.0000,-6.5421E+00)
(3.1875,-6.5721E+00)
(3.3750,-6.5665E+00)
(3.5625,-6.5348E+00)
(3.7500,-6.4837E+00)};

\end{axis}
\end{tikzpicture}&
\begin{tikzpicture}
\begin{axis}[baseline, trim axis left, title={Hencky strain, $[075](3\bar 57)$}, legend pos = outer north east,
width=5.cm, height=5.cm]
\addplot coordinates{
(0.1875,-3.7067E-01)
(0.3750,-7.4110E-01)
(0.5625,-1.1111E+00)
(0.7500,-1.4807E+00)
(0.9375,-1.8499E+00)
(1.1250,-2.2196E+00)
(1.3125,-2.5938E+00)
(1.5000,-2.9943E+00)
(1.6875,-3.6139E+00)
(1.8750,-4.8357E+00)
(2.0625,-6.1982E+00)
(2.2500,-7.4762E+00)
(2.4375,-8.6445E+00)
(2.6250,-9.7089E+00)
(2.8125,-1.0680E+01)
(3.0000,-1.1570E+01)
(3.1875,-1.2389E+01)
(3.3750,-1.3145E+01)
(3.5625,-1.3847E+01)
(3.7500,-1.4500E+01)};                
                     
\addplot coordinates{
(0.1875, 4.7010E-01)
(0.3750, 9.3008E-01)
(0.5625, 1.3802E+00)
(0.7500, 1.8206E+00)
(0.9375, 2.2518E+00)
(1.1250, 2.6747E+00)
(1.3125, 3.0915E+00)
(1.5000, 3.5141E+00)
(1.6875, 4.0355E+00)
(1.8750, 4.8146E+00)
(2.0625, 5.6132E+00)
(2.2500, 6.3357E+00)
(2.4375, 6.9809E+00)
(2.6250, 7.5591E+00)
(2.8125, 8.0809E+00)
(3.0000, 8.5554E+00)
(3.1875, 8.9902E+00)
(3.3750, 9.3913E+00)
(3.5625, 9.7639E+00)
(3.7500, 1.0112E+01)};              
                    
\addplot coordinates{
(0.1875, 1.1568E-02) 
(0.3750, 3.0805E-03) 
(0.5625,-2.0985E-02) 
(0.7500,-6.1531E-02) 
(0.9375,-1.2450E-01) 
(1.1250,-2.2522E-01) 
(1.3125,-4.0547E-01) 
(1.5000,-8.1213E-01) 
(1.6875,-2.0362E+00) 
(1.8750,-3.8631E+00) 
(2.0625,-5.0616E+00) 
(2.2500,-5.8061E+00) 
(2.4375,-6.2867E+00) 
(2.6250,-6.6030E+00) 
(2.8125,-6.8117E+00) 
(3.0000,-6.9469E+00) 
(3.1875,-7.0307E+00) 
(3.3750,-7.0778E+00) 
(3.5625,-7.0982E+00) 
(3.7500,-7.0992E+00)};
       
\addplot coordinates{
(0.1875,-3.7160E-01)
(0.3750,-7.4481E-01)
(0.5625,-1.1194E+00)
(0.7500,-1.4954E+00)
(0.9375,-1.8729E+00)
(1.1250,-2.2532E+00)
(1.3125,-2.6413E+00)
(1.5000,-3.0663E+00)
(1.6875,-3.7479E+00)
(1.8750,-4.9894E+00)
(2.0625,-6.3470E+00)
(2.2500,-7.6223E+00)
(2.4375,-8.7894E+00)
(2.6250,-9.8534E+00)
(2.8125,-1.0825E+01)
(3.0000,-1.1715E+01)
(3.1875,-1.2534E+01)
(3.3750,-1.3291E+01)
(3.5625,-1.3993E+01)
(3.7500,-1.4647E+01)};
       
\addplot coordinates{
(0.1875, 4.7085E-01)
(0.3750, 9.3300E-01)
(0.5625, 1.3866E+00)
(0.7500, 1.8318E+00)
(0.9375, 2.2690E+00)
(1.1250, 2.6992E+00)
(1.3125, 3.1253E+00)
(1.5000, 3.5626E+00)
(1.6875, 4.1156E+00)
(1.8750, 4.9009E+00)
(2.0625, 5.6928E+00)
(2.2500, 6.4097E+00)
(2.4375, 7.0505E+00)
(2.6250, 7.6251E+00)
(2.8125, 8.1441E+00)
(3.0000, 8.6164E+00)
(3.1875, 9.0497E+00)
(3.3750, 9.4499E+00)
(3.5625, 9.8219E+00)
(3.7500, 1.0170E+01)};
       
\addplot coordinates{
(0.1875,+1.2045E-02)
(0.3750,+4.3883E-03)
(0.5625,-1.9184E-02)
(0.7500,-6.0315E-02)
(0.9375,-1.2609E-01)
(1.1250,-2.3417E-01)
(1.3125,-4.3227E-01)
(1.5000,-8.8521E-01)
(1.6875,-2.1471E+00)
(1.8750,-3.8181E+00)
(2.0625,-4.9087E+00)
(2.2500,-5.5821E+00)
(2.4375,-6.0076E+00)
(2.6250,-6.2778E+00)
(2.8125,-6.4458E+00)
(3.0000,-6.5443E+00)
(3.1875,-6.5943E+00)
(3.3750,-6.6099E+00)
(3.5625,-6.6008E+00)
(3.7500,-6.5736E+00)};

\legend{$u_x$,$u_y$,$u_z$,$U_x$,$U_y$,$U_z$} 
\end{axis}
\end{tikzpicture}
\\
\begin{tikzpicture}
\begin{axis}[baseline, trim axis left, width=5.cm, height=5.cm, xlabel= Force, ylabel= $\Delta$ Displacements]
\addplot coordinates{
(0.1875, -3.7104E-01+3.7159E-01)
(0.3750, -7.4247E-01+7.4485E-01)
(0.5625, -1.1139E+00+1.1197E+00)
(0.7500, -1.4852E+00+1.4963E+00)
(0.9375, -1.8565E+00+1.8749E+00)
(1.1250, -2.2286E+00+2.2569E+00)
(1.3125, -2.6071E+00+2.6475E+00)
(1.5000, -3.0302E+00+3.0771E+00)
(1.6875, -3.8338E+00+3.7835E+00)
(1.8750, -5.1801E+00+5.0759E+00)
(2.0625, -6.5388E+00+6.4801E+00)
(2.2500, -7.7926E+00+7.7992E+00)
(2.4375, -8.9349E+00+9.0087E+00)
(2.6250, -9.9757E+00+1.0114E+01)
(2.8125, -1.0927E+01+1.1126E+01)
(3.0000, -1.1800E+01+1.2056E+01)
(3.1875, -1.2605E+01+1.2915E+01)
(3.3750, -1.3351E+01+1.3712E+01)
(3.5625, -1.4043E+01+1.4455E+01)
(3.7500, -1.4690E+01+1.5150E+01)};

\addplot coordinates{
(0.1875, 4.6919E-01-4.7083E-01)
(0.3750, 9.2646E-01-9.3307E-01)
(0.5625, 1.3721E+00-1.3870E+00)
(0.7500, 1.8063E+00-1.8330E+00)
(0.9375, 2.2296E+00-2.2716E+00)
(1.1250, 2.6432E+00-2.7038E+00)
(1.3125, 3.0502E+00-3.1330E+00)
(1.5000, 3.4711E+00-3.5750E+00)
(1.6875, 4.0668E+00-4.1430E+00)
(1.8750, 4.8832E+00-4.9554E+00)
(2.0625, 5.6589E+00-5.7717E+00)
(2.2500, 6.3507E+00-6.5115E+00)
(2.4375, 6.9657E+00-7.1744E+00)
(2.6250, 7.5159E+00-7.7707E+00)
(2.8125, 8.0119E+00-8.3114E+00)
(3.0000, 8.4627E+00-8.8058E+00)
(3.1875, 8.8756E+00-9.2616E+00)
(3.3750, 9.2565E+00-9.6851E+00)
(3.5625, 9.6100E+00-1.0081E+01)
(3.7500, 9.9401E+00-1.0455E+01)};

\addplot coordinates{
(0.1875,  1.1556E-02-1.2048E-02)
(0.3750,  3.0172E-03-4.3945E-03)
(0.5625, -2.1228E-02+1.9161E-02)
(0.7500, -6.2343E-02+6.0230E-02)
(0.9375, -1.2705E-01+1.2587E-01)
(1.1250, -2.3339E-01+2.3384E-01)
(1.3125, -4.3481E-01+4.3263E-01)
(1.5000, -9.4838E-01+8.9371E-01)
(1.6875, -2.4837E+00+2.2033E+00)
(1.8750, -4.0756E+00+3.8982E+00)
(2.0625, -5.0323E+00+4.9838E+00)
(2.2500, -5.6214E+00+5.6461E+00)
(2.4375, -5.9964E+00+6.0577E+00)
(2.6250, -6.2372E+00+6.3118E+00)
(2.8125, -6.3898E+00+6.4623E+00)
(3.0000, -6.4821E+00+6.5421E+00)
(3.1875, -6.5325E+00+6.5721E+00)
(3.3750, -6.5530E+00+6.5665E+00)
(3.5625, -6.5519E+00+6.5348E+00)
(3.7500, -6.5352E+00+6.4837E+00)};

\end{axis}
\end{tikzpicture}&
\begin{tikzpicture}
\begin{axis}[baseline, trim axis left, legend pos = outer north east, width=5.cm, height=5.cm, xlabel= Force]
\addplot coordinates{
(0.1875, -3.7067E-01+3.7160E-01)
(0.3750, -7.4110E-01+7.4481E-01)
(0.5625, -1.1111E+00+1.1194E+00)
(0.7500, -1.4807E+00+1.4954E+00)
(0.9375, -1.8499E+00+1.8729E+00)
(1.1250, -2.2196E+00+2.2532E+00)
(1.3125, -2.5938E+00+2.6413E+00)
(1.5000, -2.9943E+00+3.0663E+00)
(1.6875, -3.6139E+00+3.7479E+00)
(1.8750, -4.8357E+00+4.9894E+00)
(2.0625, -6.1982E+00+6.3470E+00)
(2.2500, -7.4762E+00+7.6223E+00)
(2.4375, -8.6445E+00+8.7894E+00)
(2.6250, -9.7089E+00+9.8534E+00)
(2.8125, -1.0680E+01+1.0825E+01)
(3.0000, -1.1570E+01+1.1715E+01)
(3.1875, -1.2389E+01+1.2534E+01)
(3.3750, -1.3145E+01+1.3291E+01)
(3.5625, -1.3847E+01+1.3993E+01)
(3.7500, -1.4500E+01+1.4647E+01)};

\addplot coordinates{
(0.1875,  4.7010E-01-4.7085E-01)
(0.3750,  9.3008E-01-9.3300E-01)
(0.5625,  1.3802E+00-1.3866E+00)
(0.7500,  1.8206E+00-1.8318E+00)
(0.9375,  2.2518E+00-2.2690E+00)
(1.1250,  2.6747E+00-2.6992E+00)
(1.3125,  3.0915E+00-3.1253E+00)
(1.5000,  3.5141E+00-3.5626E+00)
(1.6875,  4.0355E+00-4.1156E+00)
(1.8750,  4.8146E+00-4.9009E+00)
(2.0625,  5.6132E+00-5.6928E+00)
(2.2500,  6.3357E+00-6.4097E+00)
(2.4375,  6.9809E+00-7.0505E+00)
(2.6250,  7.5591E+00-7.6251E+00)
(2.8125,  8.0809E+00-8.1441E+00)
(3.0000,  8.5554E+00-8.6164E+00)
(3.1875,  8.9902E+00-9.0497E+00)
(3.3750,  9.3913E+00-9.4499E+00)
(3.5625,  9.7639E+00-9.8219E+00)
(3.7500,  1.0112E+01-1.0170E+01)};

\addplot coordinates{
(0.1875,  1.1568E-02-1.2045E-02)
(0.3750,  3.0805E-03-4.3883E-03)
(0.5625, -2.0985E-02+1.9184E-02)
(0.7500, -6.1531E-02+6.0315E-02)
(0.9375, -1.2450E-01+1.2609E-01)
(1.1250, -2.2522E-01+2.3417E-01)
(1.3125, -4.0547E-01+4.3227E-01)
(1.5000, -8.1213E-01+8.8521E-01)
(1.6875, -2.0362E+00+2.1471E+00)
(1.8750, -3.8631E+00+3.8181E+00)
(2.0625, -5.0616E+00+4.9087E+00)
(2.2500, -5.8061E+00+5.5821E+00)
(2.4375, -6.2867E+00+6.0076E+00)
(2.6250, -6.6030E+00+6.2778E+00)
(2.8125, -6.8117E+00+6.4458E+00)
(3.0000, -6.9469E+00+6.5443E+00)
(3.1875, -7.0307E+00+6.5943E+00)
(3.3750, -7.0778E+00+6.6099E+00)
(3.5625, -7.0982E+00+6.6008E+00)
(3.7500, -7.0992E+00+6.5736E+00)};

\legend{$\scriptstyle u_x-U_x$,$\scriptstyle u_y-U_y$,$\scriptstyle u_z - U_z$}
\end{axis}
\end{tikzpicture}
\end{tabular}
\end{center}
\caption{(Top) Displacements $u_x,u_y,u_z$ and $U_x,U_y,U_z$ obtained without and with use of the 3rd order elastic
  coefficients for different strain measures and different lattice orientations. (Bottom) Difference between the
  displacements of nanomembrane.\label{Cooks1}}
\end{figure}
As a FE test we consider a three dimensional variant of the Cook{'}s membrane made of 4\,nm thick silicon crystal, see
Fig.\ref{fig:parview}. The problem is analysed with and without the use of TOECs related to the Hencky and Green strain
measures, respectively; see Table \ref{3rdSi}. The membrane is discretized into $16\times16\times4$ isoparametric Lagrange
brick elements with quadratic interpolation \cite{Mazdziarz10}. The shape and boundary conditions are assumed here in the
similar way to those considered among others by \citet{MenzelS01}. The front vertical wall of membrane is conservatively
loaded by a uniformly distributed vertical traction force corresponding 3.75\,GPa. We consider two lattice orientations of
the nanomembrane, $[001](100)$ and $[075](3\bar 57)$, where the Miller indices in the round and square brackets denote
respectively the crystal orientation of nanomembrane plane and the lattice vector direction $x$ on that plane.

The resultant displacements of the mid point of the top edge of the nanomembrane are presented in Fig.\,\ref{Cooks1}. For
symmetric crystal lattice orientation $[001](100)$ the difference in displacements yielding from the use of
TOECs does not exceed 10\%, see Fig.\,\ref{Cooks1}. For asymmetric lattice orientation $[075](3\bar 57)$, the nanomembrane
bucked both for the use of the Green as well as Hencky strain measures. The use of different lattice orientation
significantly changed the behaviour of the nanomembrane, while the adding of TOECs changed its displacements up to 10\%. In
Fig.\ref{fig:parview}, the views \textit{xy} and \textit{xz} of the initial and deformed mesh for the $[075](3\bar 57)$
lattice orientation and the use of TOECs related to Hencky strain is shown.

\section{Conclusions}
In acoustic measurements, according to tradition, TOECs are determined in relation to the Green strain
\cite{HughesK53,ThurstonB64}. On the other hand, in ab-inito calculations, TOECs are often determined with the use of Biot
strain measure, e.g.\ $\varepsilon_{11}= \frac {a-a_\circ}{a_\circ}$. In result, the coefficients obtained by means of these
two methods can differ from each other dramatically. As shown in Table \ref{3rdSi} the coefficient $C_{111}= -317$\,GPa
determined experimentally with the use of Green strain is equivalent to $C_{111}= -815$\,GPa determined with the use of Biot
strain. This is because each Hookean hardens in a different way, see Fig.\,\ref{sethfig}. As a matter of fact, TOECs are
nothing more than the correction chosen for a given Hookean towards the experimentally fixed stress-stretch curve. In order
to use TOECs with another strain measure than that of used in experimental measurement, the coefficients must be
recalculated first to the given strain measure, see formula \eqref{C'} and \eqref{C111} derived by D\l u\.zewski in
\cite{Dlu00}. It is worth emphasizing that the replacement of Green strain by Hencky strain leads generally to the
decreasing of TOECs, cf.\,Table \ref{3rdSi}.

The convergence of iterative schemes applied in solving of FE boundary--value problems often depends strongly on the correct
calculation of the stiffness matrix. In this paper the analytic formulas for the tangent stiffness calculation for Hookeans
based on general Lagrangian strain have been corrected. The use of Hencky strain in FE calculations is not a new problem,
and many validation tests comparing the convergence of different algorithms were done in the past
\cite{Miehe98,MenzelS01,GermainS10}. Due to the corrections presented here, an open question arises as to how far these
errors burdened results of such validation tests in the past, cf.\ the use of \eqref{d2fTc} and their counterparts in the
concurrent numerical schemes. From the viewpoint of our computational practice, errors of this kind can limit the
convergence region and/or increase the computing time especially for the problems with expected symmetries in the resulting
stress/strain fields.

\section*{Acknowledgement}
This research was supported by the Polish Ministry of Science and Higher Education grant N\,N519\,647640.

\begin{thebibliography}{60}
\providecommand{\natexlab}[1]{#1}
\providecommand{\url}[1]{\texttt{#1}}
\expandafter\ifx\csname urlstyle\endcsname\relax
  \providecommand{\doi}[1]{doi: #1}\else
  \providecommand{\doi}{doi: \begingroup \urlstyle{rm}\Url}\fi

\bibitem[Birch(1947)]{Birch47}
F.~Birch.
\newblock Finite elastic strain of cubic crystals.
\newblock \emph{Physical Review}, 71:\penalty0 809--824, Jun 1947.
\newblock \doi{10.1103/PhysRev.71.809}.

\bibitem[Bowen and Wang(1970)]{BowenW70}
R.M.~Bowen and C.-C.~Wang.
\newblock Acceleration waves in inhomogeneous isotropic elastic bodies.
\newblock \emph{Archive of Rational Mechanics and Analysis}, 38, 1970.
\newblock \doi{10.1007/BF00251539}.
\newblock with Corrigendum in vol. 40, 403.

\bibitem[Brugger(1964)]{Brugger64}
K.~Brugger.
\newblock Thermodynamic definition of higher order elastic coefficients.
\newblock \emph{Physical Review}, 133\penalty0 (6A):\penalty0 1611--12, 1964.
\newblock \doi{10.1103/PhysRev.133.A1611}.

\bibitem[Carroll(2009)]{Carroll09}
M.M.~Carroll.
\newblock {Must Elastic Materials be Hyperelastic?}
\newblock \emph{Mathematics and Mechanics of Solids}, 14\penalty0 (4):\penalty0
  369--376, 2009.
\newblock \doi{10.1177/1081286508099385}.

\bibitem[Chadwick and Ogden(1971)]{ChadwickO71}
P.~Chadwick and R.W.~Ogden.
\newblock A theorem of tensor calculus and its application to isotropic
  elasticity.
\newblock \emph{Archive for Rational Mechanics and Analysis}, 44\penalty0
  (1):\penalty0 54--68, 1971.
\newblock ISSN 0003-9527.
\newblock \doi{10.1007/BF00250828}.

\bibitem[Chipanga(2010)]{Chipanga10}
T.~Chipanga.
\newblock \emph{Determination of the Accuracy of Residual Stress Measurement
  Methods Hole Drilling, Ultrasonic ({D}ebro-30) System and Digital
  Shearography}.
\newblock Lambert Academic Publishing, 2010.
\newblock URL \url{http://www.lap-publishing.com/}.

\bibitem[Cholewi\'nski et~al.(2014)Cholewi\'nski, Ma\'zdziarz, Jurczak, and
  D\l{}u\.zewski]{CholewinskiM14}
J.~Cholewi\'nski, M.~Ma\'zdziarz, G.~Jurczak, and P.~D\l{}u\.zewski.
\newblock Dislocation core reconstruction based on finite deformation approach
  and its application to 4{H}-{S}i{C} crystal.
\newblock \emph{International Journal for Multiscale Computational
  Engineering}, 12:\penalty0 411–421, 2014.
\newblock \doi{10.1615/IntJMultCompEng.2014010679}.

\bibitem[Clayton(2014)]{Clayton14}
J.D.~Clayton.
\newblock Analysis of shock compression of strong single crystals with
  logarithmic thermoelastic-plastic theory.
\newblock \emph{International Journal of Engineering Science}, 79\penalty0
  (0):\penalty0 1 -- 20, 2014.
\newblock ISSN 0020-7225.
\newblock \doi{10.1016/j.ijengsci.2014.02.016}.

\bibitem[Cousins(2003)]{Cousins03}
C.~S.~G.~Cousins.
\newblock Elasticity of carbon allotropes. {I}. {O}ptimization, and subsequent
  modification, of an anharmonic {K}eating model for cubic diamond.
\newblock \emph{Physical Review B}, 67:\penalty0 024107, Jan 2003.
\newblock \doi{10.1103/PhysRevB.67.024107}.

\bibitem[Curnier and Zysset(2006)]{CurnierZ06}
Z.~Curnier and Ph.~Zysset.
\newblock A family of metric strains and conjugate stresses, prolonging usual
  material laws from small to large transformations.
\newblock \emph{International Journal of Solids and Structures}, 43:\penalty0
  3057--3086, 2006.
\newblock \doi{10.1016/j.ijsolstr.2005.06.015}.

\bibitem[Deputat et~al.(1993)Deputat, Szelazek, and Adamski]{DeputatS92}
J.~Deputat, J.~Szelazek, and M.~Adamski.
\newblock \emph{Experiences in Ultrasonic Measurements of Stresses in Rails},
  pages 109--118.
\newblock Springer Netherlands, Dordrecht, 1993.
\newblock ISBN 978-94-015-8151-6.
\newblock \doi{10.1007/978-94-015-8151-6_9}.

\bibitem[Destrade and Ogden(2010)]{DestradeO10}
M.~Destrade and R.W.~Ogden.
\newblock On the third- and fourth-order constants of incompressible isotropic
  elasticity.
\newblock \emph{Journal of the Acoustical Society of America}, 128\penalty0
  (6):\penalty0 3334--3343, 2010.
\newblock \doi{10.1121/1.3505102}.

\bibitem[D\l{}u\.zewski(2000)]{Dlu00}
P.~D\l{}u\.zewski.
\newblock Anisotropic hyperelasticity based upon general strain measures.
\newblock \emph{Journal of Elasticity}, 60\penalty0 (2):\penalty0 119--129,
  2000.
\newblock \doi{10.1023/A:1010969000869}.

\bibitem[D\l{}u\.zewski and Traczykowski(2003)]{DluT03}
P.~D\l{}u\.zewski and P.~Traczykowski.
\newblock Numerical simulation of atomic positions in quantum dot by means of
  molecular statics.
\newblock \emph{Archive of Mechanics}, 55:\penalty0 501--515, 2003.

\bibitem[Egle and Bray(1969)]{EgleB69}
D.M.~Egle and D.E.~Bray.
\newblock Measurement of acoustoelastic and third-order elastic constants for
  rail steel.
\newblock \emph{J. Acoust. Soc. Am.}, 60:\penalty0 741, 1969.
\newblock \doi{10.1121/1.381146}.

\bibitem[Feng et~al.(2016)Feng, Levitas, and Hemley]{FengL16}
B.~Feng, V.I.~Levitas, and R.J.~Hemley.
\newblock Large elastoplasticity under static megabar pressures: Formulation
  and application to compression of samples in diamond anvil cells.
\newblock \emph{International Journal of Plasticity}, 84:\penalty0 33 -- 57,
  2016.
\newblock ISSN 0749-6419.
\newblock \doi{10.1016/j.ijplas.2016.04.017}.

\bibitem[Frogley et~al.(2000)Frogley, Downes, and Dunstan]{FrogleyD00}
M.D.~Frogley, J.R.~Downes, and D.J.~Dunstan.
\newblock Theory of the anomalously low band-gap pressure coefficients in
  strained-layer semiconductor alloys.
\newblock \emph{Physical Review B}, 62:\penalty0 13612--13616, Nov 2000.
\newblock \doi{10.1103/PhysRevB.62.13612}.

\bibitem[Gantmacher(1954)]{Gantmacher54}
F.F.~Gantmacher,
\newblock \emph{The theory of matrices}.
\newblock Goz. Izdat. Lit., Moscow, 1954; English transl., Applications of the
  theory of matrices, Interscience, New York, 1959, 1954.

\bibitem[Germain et~al.(2010)Germain, Scherer, and Steinmann]{GermainS10}
S.~Germain, M.~Scherer, and P.~Steinmann.
\newblock On inverse form finding for anisotropic hyperelasticity in
  logarithmic strain space.
\newblock \emph{International Journals of Structural Changes in Solids –--
  Mechanics and Applications}, 2\penalty0 (2):\penalty0 1--16, 2010.

\bibitem[Hankey and Schuele(1970)]{HankeyS70}
R.E.~Hankey and D.E.~Schuele.
\newblock Third-order elastic constants of {A}l$_2${O}$_3$.
\newblock \emph{Journal of the Acoustical Society of America}, 48\penalty0
  (1B):\penalty0 190--202, 1970.
\newblock \doi{10.1121/1.1912116}.

\bibitem[Higham(2008)]{Higham08}
N.J.~Higham.
\newblock \emph{Functions of Matrices: {Theory} and Computation}.
\newblock Society for Industrial and Applied Mathematics, Philadelphia, PA,
  USA, 2008.
\newblock ISBN 978-0-898716-46-7.

\bibitem[Hiki and Granato(1966)]{HikiG66}
Y.~Hiki and A.V.~Granato.
\newblock Anharmonicity in noble metals; higher order elastic constants.
\newblock \emph{Physical Review}, 144\penalty0 (2):\penalty0 411--19, 1966.
\newblock \doi{10.1103/PhysRev.144.411}.

\bibitem[Hill(1970)]{Hill70}
R.~Hill.
\newblock Constitutive inequalities for isotropic solids under finite strain.
\newblock \emph{Proceedings of Royal Society of London A}, 314:\penalty0
  457--472, 1970.
\newblock \doi{10.1098/rspa.1970.0018}.

\bibitem[Hill(1978)]{Hill78}
R.~Hill.
\newblock Aspects of invariance in solid mechanics.
\newblock \emph{Advances in Applied Mechanics}, 18:\penalty0 1--75, 1978.
\newblock \doi{10.1016/S0065-2156(08)70264-3}.

\bibitem[Hirth and Lothe(1982)]{HirthL82}
J.P.~Hirth and J.~Lothe.
\newblock \emph{Theory of Dislocations}.
\newblock Wiley, New York, 1982.

\bibitem[Hmiel et~al.(2016)Hmiel, Winey, Gupta, and Desjarlais]{HmielW16}
A.~Hmiel, J.~M. Winey, Y.~M. Gupta, and M.~P. Desjarlais.
\newblock Nonlinear elastic response of strong solids: First-principles
  calculations of the third-order elastic constants of diamond.
\newblock \emph{Physical Review B}, 93:\penalty0 174113, May 2016.
\newblock \doi{10.1103/PhysRevB.93.174113}.

\bibitem[Horodon and Averbach(1961)]{HorodonA61}
M.J.~Horodon and B.L.~Averbach.
\newblock Precision density measurements on deformed copper and aluminum single
  crystals.
\newblock \emph{Acta Metallurgica}, 9:\penalty0 247--249, 1961.
\newblock \doi{10.1016/0001-6160(61)90074-8}.

\bibitem[Hughes and Kelly(1953)]{HughesK53}
D.~S. Hughes and J.~L. Kelly.
\newblock Second-order elastic deformation of solids.
\newblock \emph{Physical Review}, 92:\penalty0 1145--1149, Dec 1953.
\newblock \doi{10.1103/PhysRev.92.1145}.

\bibitem[Johal and Dunstan(2006)]{JohalD06}
A.S.~Johal and D.J.~Dunstan.
\newblock Reappraisal of experimental values of third-order elastic constants
  of some cubic semiconductors and metals.
\newblock \emph{Physical Review B}, 73:\penalty0 024106, Jan 2006.
\newblock \doi{10.1103/PhysRevB.73.024106}.

\bibitem[Jones and Menon(2014)]{JonesM14}
S.~Jones and Ch.S.~Menon.
\newblock Non-linear elastic behavior of hexagonal silicon carbide.
\newblock \emph{physica status solidi (b)}, 251\penalty0 (6):\penalty0
  1186--1191, 2014.
\newblock ISSN 1521-3951.
\newblock \doi{10.1002/pssb.201451024}.

\bibitem[Jurczak and D{\l}u\.{z}ewski(2016)]{Jurczak16}
G.~Jurczak and P.~D{\l}u\.{z}ewski.
\newblock {Finite element modelling of nonlinear piezoelectricity in wurtzite
  GaN/AlN quantum dots }.
\newblock \emph{Computational Materials Science}, 111:\penalty0 197--202, 2016.
\newblock ISSN 0927-0256.
\newblock \doi{10.1016/j.commatsci.2015.09.024}.

\bibitem[Jurczak and u\.zewski(2018)]{JurczakD18}
G.~Jurczak and P.~D\l u\.zewski.
\newblock Finite element modelling of threading dislocation effect on polar
  {G}a{N}/{A}l{N} quantum dot.
\newblock \emph{Physica E: Low-dimensional Systems and Nanostructures},
  95:\penalty0 11--15, 2018.
\newblock ISSN 1386-9477.
\newblock \doi{10.1016/j.physe.2017.08.018}.

\bibitem[Lang and Gupta(2011)]{LangG11}
J.M.~Lang and Y.M.~Gupta.
\newblock Experimental determination of third-order elastic constants of
  diamond.
\newblock \emph{Physical Review Letters}, 106:\penalty0 125502, Mar 2011.
\newblock \doi{10.1103/PhysRevLett.106.125502}.

\bibitem[\L{}epkowski(2008)]{Lepkowski08}
S.P.~\L{}epkowski.
\newblock Significance of third-order elasticity for determination of the
  pressure coefficient of the light emission in strained quantum wells.
\newblock \emph{Physical Review B}, 78:\penalty0 153307, Oct 2008.
\newblock \doi{10.1103/PhysRevB.78.153307}.

\bibitem[\L{}opuszy\ifmmode~\acute{n}\else \'{n}\fi{}ski and
  Majewski(2007)]{LopuszynskiM07}
M.~\L{}opuszy\ifmmode~\acute{n}\else \'{n}\fi{}ski and J.A.~Majewski.
\newblock Ab initio calculations of third-order elastic constants and related
  properties for selected semiconductors.
\newblock \emph{Physical Review B}, 76:\penalty0 045202, Jul 2007.
\newblock \doi{10.1103/PhysRevB.76.045202}.

\bibitem[Ma\'zdziarz(2010)]{Mazdziarz10}
M.~Ma\'zdziarz.
\newblock Unified isoparametric {3D} lagrange finite elements.
\newblock \emph{Computer Modeling in Engineering \& Sciences}, 66\penalty0
  (1):\penalty0 1--24, 2010.
\newblock \doi{10.3970/cmes.2010.066.001}.

\bibitem[Ma\'zdziarz and Gajewski(2015)]{Mazdziarz15}
M.~Ma\'zdziarz and M.~Gajewski.
\newblock Estimation of isotropic hyperelasticity constitutive models to
  approximate the atomistic simulation data for aluminium and tungsten
  monocrystals.
\newblock \emph{Computer Modeling in Engineering \& Sciences}, 105\penalty0
  (2):\penalty0 123--150, 2015.
\newblock \doi{10.3970/cmes.2015.105.123}.

\bibitem[McSkimin and Jr.(1972)]{McSkiminA72}
H.J.~McSkimin and P.~Andreatch Jr.
\newblock Elastic moduli of diamond as a function of pressure and temperature.
\newblock \emph{Journal of Applied Physics}, 43\penalty0 (7):\penalty0
  2944--2948, 1972.
\newblock \doi{10.1063/1.1661636}.

\bibitem[Menzel and Steinmann(2001)]{MenzelS01}
A.~Menzel and P.~Steinmann.
\newblock On the comparison of two strategies to formulate orthotropic
  hyperelasticity.
\newblock \emph{Journal of Elasticity}, 62:\penalty0 171–201, 2001.
\newblock \doi{10.1023/A:101293750}.

\bibitem[Miehe(1998)]{Miehe98}
C.~Miehe.
\newblock Comparison of two algorithms for the computation of fourth-order
  isotropic tensor functions.
\newblock \emph{Computers \& Structures}, 66:\penalty0 37--l3, 1998.
\newblock \doi{10.1016/S0045-7949(97)00073-4}.

\bibitem[Murnaghan(1937)]{Murnaghan37}
F.~D.~Murnaghan.
\newblock Finite deformations of an elastic solid.
\newblock \emph{American Journal of Mathematics}, 59\penalty0 (2):\penalty0
  235--260, 1937.
\newblock ISSN 00029327.
\newblock \doi{10.2307/2371405}.

\bibitem[Nielsen(1986)]{Nielsen86}
O.~H.~Nielsen.
\newblock Optical phonons and elasticity of diamond at megabar stresses.
\newblock \emph{Physical Review B}, 34:\penalty0 5808--5819, Oct 1986.
\newblock \doi{10.1103/PhysRevB.34.5808}.

\bibitem[Norris(2008)]{Norris08}
Andrew~N.~Norris.
\newblock Higher derivatives and the inverse derivative of a tensor-valued
  function of a tensor.
\newblock \emph{Quart. Appl. Math.}, 66\penalty0 (4):\penalty0 725--741, 2008.
\newblock \doi{10.1090/S0033-569X-08-01108-2}.

\bibitem[Ogden(1984)]{Ogden84}
R.~W. Ogden.
\newblock \emph{Non-Linear Elastic Deformations}.
\newblock Ellis Horwood Ltd., Chichester, 1984.

\bibitem[Poirier and Tarantola(1998)]{Poirier98}
J.-P~Poirier and A.~Tarantola.
\newblock A logarithmic equation of state.
\newblock \emph{Physics of the Earth and Planetary Interiors}, 109\penalty0
  (1–2):\penalty0 1 -- 8, 1998.
\newblock ISSN 0031-9201.
\newblock \doi{10.1016/S0031-9201(98)00112-5}.

\bibitem[Schramm(1999)]{Schramm99}
R.~E.~Schramm.
\newblock Ultrasonic measurement of stress in railroad wheels.
\newblock \emph{Review of Scientific Instruments}, 70\penalty0 (2):\penalty0
  1468--1472, 1999.
\newblock \doi{10.1063/1.1149607}.

\bibitem[Seth(1964)]{Seth62}
B.~R. Seth.
\newblock Generalized strain measure with applications to physical problems.
\newblock In M.~Reiner and D.~Abir, editors, \emph{Second-Order Effects in
  Elasticity, Plasticity and Fluid Dynamics, Proceedings of International
  Symposium, Haifa, April 23-27, 1962}, pages 162--172, Oxford, 1964. Pergamon
  Press.

\bibitem[Singh et~al.(2011)Singh, Pandey, Singh, and Yadav]{SinghP11}
D.~Singh, D.K.~Pandey, D.K.~Singh, and R.R.~Yadav.
\newblock Propagation of ultrasonic waves in neptunium monochalcogenides.
\newblock \emph{Applied Acoustics}, 72\penalty0 (10):\penalty0 737 -- 741,
  2011.
\newblock ISSN 0003-682X.
\newblock \doi{10.1016/j.apacoust.2011.04.002}.

\bibitem[Singh et~al.(2008)Singh, Singh, and Singh]{SinghS08}
R.~K. Singh, R.~P. Singh, and M.~P. Singh.
\newblock Acoustical characterization of nanostructured metals.
\newblock \emph{International Journal of Nanoscience}, 07\penalty0
  (06):\penalty0 315--323, 2008.
\newblock \doi{10.1142/S0219581X08005481}.

\bibitem[Spaepen(2000)]{Spaepen00}
F.~Spaepen.
\newblock Interfaces and stresses in thin films.
\newblock \emph{Acta Materialia}, 48:\penalty0 31--42, 2000.
\newblock \doi{10.1016/S1359-6454(99)00286-4}.

\bibitem[Thurston and Brugger(1964)]{ThurstonB64}
R.~N.~Thurston and K.~Brugger.
\newblock Third-order elastic constants and the velocity of small amplitude
  elastic waves in homogeneouly stressed media.
\newblock \emph{Physical Review}, 133\penalty0 (6A):\penalty0 1604--1610, 1964.
\newblock \doi{10.1103/PhysRev.135.AB3.2}.

\bibitem[Thurston et~al.(1966)Thurston, McSkimin, and Andreatch]{ThurstonS66}
R.~N.~Thurston, H.~J.~McSkimin, and P.~Andreatch.
\newblock Third--order elastic coefficients of quartz.
\newblock \emph{Journal of Applied Physics}, 37\penalty0 (1):\penalty0
  267--275, 1966.
\newblock \doi{10.1063/1.1707824}.

\bibitem[Walker et~al.(1985)Walker, Saunders, and Hawkey]{Walker85}
N.~J.~Walker, G.~A.~Saunders, and J.~E.~Hawkey.
\newblock Soft {TA} models and anharmonicity in cadmium telluride.
\newblock \emph{Philosophical Magazine B}, 52\penalty0 (5):\penalty0
  1005--1018, 1985.
\newblock \doi{10.1080/01418638508241890}.

\bibitem[Wallace(1970)]{Wallace70}
Duane~C.~Wallace.
\newblock Thermoelastic theory of stressed crystals and higher-order elastic
  constants.
\newblock \emph{Solid State Physics}, 25:\penalty0 301 -- 404, 1970.
\newblock ISSN 0081-1947.
\newblock \doi{10.1016/S0081-1947(08)60010-7}.

\bibitem[Wang and Li(2009)]{WangL09}
Hao~Wang and Mo~Li.
\newblock {\em Ab initio} calculations of second-, third-, and fourth-order
  elastic constants for single crystals.
\newblock \emph{Physical Review B}, 79:\penalty0 224102, Jun 2009.
\newblock \doi{10.1103/PhysRevB.79.224102}.

\bibitem[Wang et~al.(2010)Wang, Wang, Wu, Yao, and Liu]{WangW10}
Rui Wang, Shaofeng Wang, Xiaozhi Wu, Yin Yao, and Anping Liu.
\newblock Ab initio calculations on the third-order elastic constants for
  selected {B}2–{M}g{RE} ({RE} = {Y}, {T}b, {D}y, {N}d) intermetallics.
\newblock \emph{Intermetallics}, 18\penalty0 (12):\penalty0 2472 -- 2476, 2010.
\newblock ISSN 0966-9795.
\newblock \doi{10.1016/j.intermet.2010.08.039}.

\bibitem[Wang et~al.(2013)Wang, Gu, Sun, and Zhang]{WangG13}
X.~Wang, Y.~Gu, Xu~Sun, and Yue Zhang.
\newblock Nonlinear elastic response of cubic crystals to biaxial strain.
\newblock \emph{Computational Materials Science}, 79\penalty0 (0):\penalty0 284
  -- 288, 2013.
\newblock ISSN 0927-0256.
\newblock \doi{10.1016/j.commatsci.2013.05.058}.

\bibitem[Winey et~al.(2016)Winey, Hmiel, and Gupta]{WineyH16}
J.M.~Winey, A.~Hmiel, and Y.M.~Gupta.
\newblock Third-order elastic constants of diamond determined from experimental
  data.
\newblock \emph{Journal of Physics and Chemistry of Solids}, 93:\penalty0
  118--120, 2016.
\newblock ISSN 0022-3697.
\newblock \doi{10.1016/j.jpcs.2016.02.016}.

\bibitem[Young et~al.(2007)Young, Kioseoglou, Dimitrakopulos, D\l{}u\.zewski,
  and Komninou]{YoungK07}
T.D~Young, J.~Kioseoglou, G.P.~Dimitrakopulos, P.~D\l{}u\.zewski, and Ph.~Komninou.
\newblock {3D} modelling of misfit networks in the interface region of
  heterostructures.
\newblock \emph{Journal of Physics D: Applied Physics}, 40\penalty0
  (13):\penalty0 4084, 2007.
\newblock \doi{10.1088/0022-3727/40/13/027}.

\bibitem[Zhao et~al.(2007)Zhao, Winey, and Gupta]{ZhaoW07}
J.~Zhao, J.M.~Winey, and Y.M.~Gupta.
\newblock First-principles calculations of second- and third-order elastic
  constants for single crystals of arbitrary symmetry.
\newblock \emph{Physical Review B}, 75:\penalty0 094105, Mar 2007.
\newblock \doi{10.1103/PhysRevB.75.094105}.

\end{thebibliography}

\appendix 
\section{Appendix}
For isoparametric FEs the substitution of \eqref{Pmi} into
${\mathbf K}_{\sf mn} \stackrel{df}{=} \fracp{{\mathbf P}_{\sf m}}{{\mathbf a}_{\sf n}}$ can be rewritten in the following form
\begin{align}
K{_{\sf m}}^i{_{\sf n}}^j 
 &= \int_{V} \Big(
\fracp{\widehat{\varepsilon}_{PR}}{a_{{\sf m}i}}
\fracp{\widehat{\sigma}^{PR}}{\widehat{\varepsilon}_{ST}}
\fracp{\widehat{\varepsilon}_{ST}}{a_{{\sf n}j}} + {\widehat{\sigma}^{PR}} \fracp{^2\widehat{\varepsilon}_{PR}}{a_{{\sf m}i}\partial a_{{\sf n}j}}
 \Big) \det{\mathbf F}^{-1} dv \label{aKminj}\ .
\end{align}
$\fracp{^2\widehat{\boldsymbol\varepsilon}} {{\mathbf a}_{\sf m}\partial {\mathbf a}_{\sf n}}$ can be considered as the second
derivative of a complex tensor function $\widehat{\boldsymbol\varepsilon}=\widehat{\boldsymbol\varepsilon} ({\mathbf E})$, where
the Green strain depends on nodal displacement vectors, ${\mathbf E}={\mathbf E}({\mathbf a}_{\sf 1},\ldots,{\mathbf a}_{\sf
  k})$. According to the chain rule this derivative reads
\begin{align}
  \fracp{^2\widehat{\varepsilon}_{PR}} {a_{{\sf m}i}\partial a_{{\sf n}j}} \label{deaa}
  = \fracp{\widehat{\varepsilon}_{PR}}  {E_{ST}} \fracp{^2E_{ST}}{a_{{\sf m}i}\,\partial a_{{\sf n}j}} 
  +\fracp{^2\widehat{\varepsilon}_{PR}} {E_{ST} \partial E_{MN}} \fracp{E_{ST}}{a_{{\sf m}i}} \fracp{E_{MN}}{a_{{\sf n}j}} ,  
\end{align}
where
\begin{align}
\fracp{E_{KL}}{a_{{\sf m}i}} & = \frac 1 2 \big( N_{{\sf mi},K} {{F}^i}_{L} + N_{{\sf mi},L} {{F}^i}_{K} \big),&
\fracp{^2E_{KL}}{a_{{\sf m}i}\,\partial a_{{\sf n}j}} & =  g^{ij} N_{{\sf mi},K} N_{{\sf nj},L},&
\fracp{^3E_{KL}}{a_{{\sf m}i}\,\partial a_{{\sf n}j}\,\partial a_{{\sf a}b}} & = 0\ . \label{deda12}
\end{align}
Substitution of \eqref{deda12} into \eqref{deaa}, and the following result into \eqref{aKminj} gives
subsequently
\begin{align}
K{_{\sf m}}^i{_{\sf n}}^j 
&= \int_{V} \bigg(\frac 1 2 \big({g^i}_p\, N_{{\sf mi},r} + {g^i}_r\, N_{{\sf mi},p} \big)\,{{\mathcal A}_{PR}}^{pr}\, 
    \fracp{\widehat{\sigma}^{PR}}{\widehat{\varepsilon}_{ST}}  
  {{\mathcal A}_{ST}}^{st} \frac 1 2 \big({g^j}_s\, N_{{\sf mj},t} + {g^j}_t\, N_{{\sf mj},s} \big)\nonumber\\
&\qquad + \frac 1 2 \big( N_{{\sf mi},S} {{F}^i}_{T} + N_{{\sf ni},T} {{F}^i}_{S}\big)\, \widehat{\sigma}^{PR}\fracp{^2\widehat{\varepsilon}_{PR}} {E_{ST} \partial E_{KL}} \,\frac 1 2 \big( N_{{\sf mj},K} {{F}^j}_{L} + N_{{\sf nj},L} {{F}^j}_{K}\big)\label{Kminj}\\
&\qquad\qquad\qquad\qquad\qquad\qquad\qquad\qquad\qquad\quad + g^{ij} N_{{\sf mi},S} N_{{\sf mj},T}\, \fracp{\widehat{\varepsilon}_{PR}}  {E_{ST}} \widehat{\sigma}^{PR}
\bigg) \det{\mathbf F}^{-1} dv\nonumber\\
= \int_{V}\!\!  N_{{\sf mi},k}& \Big[\Big({\mathcal A}{_{PR}}^{ik}
\fracp{\widehat{\sigma}^{PR}}{\widehat{\varepsilon}_{ST}} {{\mathcal A}_{ST}}^{jl} +  
\widehat{\sigma}^{PR}\,F^i{_S} F^k{_T}\fracp{^2\widehat{\varepsilon}_{PR}} {E_{ST} E_{MN}} F^j{_M} F^l{_N} \Big) 
\det{\mathbf F}^{-1}+ g^{ij} \sigma^{kl}\Big] N_{{\sf nj},l}\, dv\,. \nonumber
\end{align}
The last term leads to \eqref{Kij} where
${\mathcal B}{_{PR}}^{ikjl} = F{^i}_IF{^k}_KF{^j}_JF{^l}_L\,\fracp{^2\widehat{\varepsilon}_{PR}}{E_{IJ}\partial
  E_{KL}}$. According to \eqref{d2fTbd} the second derivative $\fracp{^2\widehat{\boldsymbol\varepsilon}}{{\mathbf E}^2}$
rewritten in the stretch eigenvector basis reads
\begin{align}
\fracp{^2\widehat{\varepsilon}^{i\mspace{-4.5mu}j}} {E_{j\mspace{-3.0mu}k} E_{k\mspace{-2.0mu}i}} & = \begin{cases}  
u_{\sf i}^4\, \varepsilon''(E_{\sf i})  & \text{for}\quad u_{\sf i}= u_{\sf j}= u_{\sf k} \,,\\
2u_{\sf i}u_{\sf j}^3\frac{\frac{\varepsilon(E_{\sf j})-\varepsilon(E_{\sf i})}{E_{\sf j}-E_{\sf i}} 
- \varepsilon'(E_{\sf k})}{E_{\sf j}-E_{\sf i}}  & \text{for}\quad u_{\sf i}\ne u_{\sf j} = u_{\sf k} \,,\\ 
-2u_{\sf i}u_{\sf j}u_{\sf k}^2\, \frac{\varepsilon(E_{\sf i}) ( E_{\sf j}-E_{\sf k}) + \varepsilon(E_{\sf j}) ( E_{\sf k}-E_{\sf i}) + \varepsilon(E_{\sf k}) ( E_{\sf i}-E_{\sf j} )}
     {(E_{\sf i}-E_{\sf j})(E_{\sf j}-E_{\sf k})( E_{\sf k}-E_{\sf i})}
  & \text{for}\quad u_{\sf i}\ne u_{\sf j}\ne u_{\sf k}\ne  u_{\sf i}\,, \label{Bijjkki} 
\end{cases} 
\end{align} 
where 
\begin{subequations}
\begin{align}
E_{\sf i} &= \frac12 (u_{\sf i}^2-1)\,,\\
\widehat{\varepsilon}(E_{\sf i}) &= f(u_{\sf i})\,,\\
\widehat{\varepsilon}\,'(E_{\sf i})&= \fracp{f\big(u_{\sf i}(E_{\sf i})\big)}{E_{\sf i}} 
= \fracp f{u_{\sf i}} \fracp{u_{\sf i}}{E_{\sf i}} = u_{\sf i}^{-1} f'(u_{\sf i})\,,\\
\widehat{\varepsilon}\,''(E_{\sf i}) &= \fracp{^2 f}{E_{\sf i}^2} = \fracp \ {E_{\sf i}} \bigg( \fracp f {E_{\sf i}} \bigg) 
= \fracp \ {u_{\sf i}} \bigg( \fracp f {u_{\sf i}} \fracp{u_{\sf i}}{E_{\sf i}} \bigg) 
\fracp{u_{\sf i}}{E_{\sf i}} =  u_{\sf i}^{-2} f''(u_{\sf i}) - u_{\sf i}^{-3} f'(u_{\sf i})\,.
\end{align}\label{E'''}
\end{subequations} 
\end{document}